\documentclass[amsmath,amssymb,11pt,tightenlines,superscriptaddress,nofootinbib,preprintnumbers, notitlepage,aps,prd,onecolumn,]{revtex4-1}

\usepackage[utf8]{inputenc}
\usepackage[english]{babel}
\usepackage{natbib}
\usepackage{multirow}
\usepackage{amsmath,amssymb}
\usepackage{graphicx}
\usepackage{caption}
\usepackage{subcaption}
\usepackage{color}
\usepackage{url}
\usepackage{slashed}
\usepackage{footmisc}
\usepackage{hyperref}
\usepackage{ragged2e}
\DeclareCaptionJustification{justified}{\justifying}
\newcommand{\be}{\begin{equation}}
\newcommand{\ee}{\end{equation}}
\newcommand{\bea}{\begin{eqnarray}}
\newcommand{\eea}{\end{eqnarray}}

\usepackage{natbib}
\usepackage{graphicx}

\begin{document}

\title{Di-Higgs Blind Spots in Gravitational Wave Signals}

\author{Alexandre Alves} 
\affiliation{Departamento de F\'{i}sica, Universidade Federal de S\~{a}o Paulo, UNIFESP, Diadema, 09972-270, Brazil}
\author{Dorival Gon\c{c}alves} 
\affiliation{Department of Physics, Oklahoma State University, Stillwater, OK, 74078, USA}
\author{Tathagata Ghosh}
\affiliation{Department of Physics \& Astronomy, University of Hawaii, Honolulu, HI 96822, USA}
\affiliation{PITT PACC, Department of Physics and Astronomy, University of Pittsburgh, 3941 O'Hara St., Pittsburgh, PA 15260, USA}
\author{Huai-Ke Guo}
\author{Kuver Sinha}
\affiliation{Department of Physics and Astronomy, University of Oklahoma, Norman, OK 73019, USA}

\preprint{OSU-HEP-20-09, PITT-PACC 2004}

\begin{abstract}Conditions for strong first-order phase transition and  generation of observable gravitational wave (GW) signals are very restrictive to the profile of the Higgs potential. Working in the minimal extension of the SM with a new gauge singlet real scalar, we show that the production of signals relevant for  future GW experiments, such as LISA, can favor  depleted resonant and non-resonant di-Higgs rates at colliders  for  phenomenologically relevant regimes of scalar mixing angles and masses for the heavy scalar. We  perform a comprehensive study on the emergence of these di-Higgs blind spot configurations in GWs and also show that di-boson channels, $ZZ$ and $WW$, can restore the phenomenological complementarities between GW and collider experiments in these parameter space regimes.\end{abstract}

\maketitle


\section{Introduction}

The Higgs potential is a vital element of the Standard Model (SM) of particle physics. Its measurement is crucial for understanding the exact mechanism of electroweak symmetry breaking and the origin of mass in our universe~\cite{Higgs:1964ia,PhysRevLett.13.508,PhysRevLett.13.321}. Distinct shapes for the Higgs potential can display contrasting patterns of electroweak symmetry breaking in the early universe, from a smooth crossover in the SM to  a strong first-order phase transition with  new physics contributions~\cite{Kajantie:1996mn,Kajantie:1996qd,Grojean:2004xa,Kanemura:2004ch,Noble:2007kk,Huang:2015tdv,Kobakhidze:2015xlz,Chen:2017qcz,Chao:2017vrq,Chao:2017ilw,Jain:2017sqm,deVries:2017ncy,Reichert:2017puo,Bian:2017wfv,Carena:2018vpt,Zhou:2020idp}. 

 Higgs pair production $pp\rightarrow hh$ provides a direct probe of the Higgs potential at colliders~\cite{DiMicco:2019ngk, Eboli:1987dy,Plehn:1996wb,Abramowicz:2016zbo,Roloff:2019crr}. This process is of central importance in measuring  the triple Higgs coupling as well as new heavy scalar interactions in the Higgs sector via non-resonant and resonant di-Higgs searches, respectively. Current ATLAS and CMS high-luminosity projections indicate that the triple Higgs coupling will be constrained in the range ${ 0.1<\lambda_3/\lambda_{3,\mathrm{SM}}<2.3}$ at 95\% CL~\cite{Cepeda:2019klc}. Resonant searches are also  being performed resulting in significant  limits~\cite{Sirunyan:2019quj}. For the latter, the weak boson fusion process provides relevant additional new physics sensitivity~\cite{Barman:2020ulr}. The measurement of the Higgs potential, in particular the Higgs self-interactions, will remain as one of the prime targets for the Large Hadron Collider (LHC) and provides a strong motivation for future colliders~\cite{Barr:2014sga,Azatov:2015oxa,He:2015spf,Fuks:2015hna,Dawson:2015oha,Buttazzo:2015bka,Banerjee:2018yxy,Goncalves:2018qas,Biekotter:2018jzu,Kim:2019wns}.

Gravitational Wave (GW) experiments, such as the future Laser Interferometer Space Antenna (LISA)~\cite{amaroseoane2017laser},  
Big Bang Observer (BBO)~\cite{Crowder:2005nr},  DECi-hertz Interferometer Gravitational wave 
Observatory (DECIGO)~\cite{Yagi:2011wg}, Taiji~\cite{Gong:2014mca}, and Tianqin~\cite{Luo:2015ght}, present a new window to access the Higgs potential. First-order phase transitions, that arise from  a scalar field tunneling from a local to a true minimum across an energy barrier, result in a relevant source of gravitational radiation. In general, the significant characteristics of the effective potential  are the relative depth of the true minimum, the height of the  barrier that separates the true minimum from the false one, and the distance between the two minima in field space at the nucleation temperature~\cite{Chala:2019rfk}. While this is an apparent simple picture, it has  interesting phenomenological implications when the Higgs boson mixes with other scalars. This is because the required conditions for experimentally detectable GW signals are in general very restrictive to the shape of the Higgs potential~\cite{Espinosa:2010hh,Espinosa:2011ax}.

Working in the minimal extension of the SM with a new gauge singlet real scalar, commonly known in the literature as ``xSM"~\cite{Profumo:2007wc,Profumo:2014opa,Chen:2014ask,Huang:2017jws,Alves:2018jsw, Alves:2017ued, Alves:2018oct, Alves:2019igs}, we show in this paper that the conditions for obtaining large GW signals from a  first-order phase transition can favor suppressed  branching ratios of the heavy scalar $h_2$ to di-higgs $h_2\rightarrow h_1h_1$ in specific $m_{h_2}$ regimes,  even above the Higgs boson threshold and with relatively large mixing angles. The same parameter regime displays characteristic Higgs self-couplings with  suppressed non-resonant di-Higgs cross-sections.
These observations have significant consequences for the complementarity of probes for the Higgs sector using GW and collider experiments. They would imply that while LISA would be sensitive to  certain parameter space regions, the LHC would not be able to observe the corresponding regions via di-Higgs resonant or non-resonant production. We dub these phenomenologically important parameter space regions as \emph{di-Higgs blind spots}. 

We  carefully study the phenomenological conditions on the Higgs potential, as well as the parameters governing the observation of GWs, that restrict the Higgs sector to the blind spots. This  includes a detailed exploration of the shape of the Higgs potential, in the vicinity of these parameter space regimes, through scrutiny of the Higgs couplings,  potential barrier, potential  depth, and the separation of the minima during the phase transition. We  find that whereas the di-Higgs channel cannot lead to complementary LHC signals, the collider reciprocity can nonetheless be restored with other relevant decay channels: $h_2\rightarrow WW, ZZ$. We go on to  perform  detailed analyses of di-Higgs and di-boson searches at blind spot benchmarks, showing their phenomenological complementarity to GW studies.\footnote{It is important to highlight that the observation of GW signals can only favor feeble $\mathcal{BR}(h_2\rightarrow h_1h_1)$ for small or intermediary $m_{h_2}$, in respect to the EW scale. The decays of $h_2$ to vector bosons are fully determined by their Goldstone nature for $m_{h_2}\gg m_W$, where $\mathcal{BR}(h_2\rightarrow h_1h_1)=\mathcal{BR}(h_2\rightarrow ZZ)=\mathcal{BR}(h_2\rightarrow WW)/2=1/4$.}

This paper is structured as follows. In Sec.~\ref{sec:blindspot}, we show the emergence of blind spots in di-Higgs production at the LHC assuming the xSM model. Next, we study the Higgs potential in the vicinity of the blind spot and discuss the sensitivity prospects to gravitational wave signals. We pay particular attention to the parameters that control the stochastic gravitational wave signals. This singles out the appearance of these suppressed heavy scalar branching ratio regions. In Sec.~\ref{sec:collider}, we perform a collider analysis  using the di-Higgs and di-boson channels. Finally, we present a summary in Sec.~\ref{sec:summary}.

\section{\label{sec:blindspot}Di-Higgs Blind Spots}

In this Section, we build up our discussion in three stages. We first provide a short summary with the general features of the xSM model, then show that blind spots can appear simultaneously in resonant as well as non-resonant di-Higgs production at colliders. Finally,  we study the scalar potential in the vicinity of these relevant parameter space regions, paying particular attention to the behavior of parameters that control the stochastic gravitational wave signals.

\subsection{Scalar Potential}

We consider the extension of the SM where there is an additional SM gauge singlet real scalar field~\cite{Profumo:2007wc}
\begin{eqnarray}
  V(H,S) &=& -\mu^2 H^{\dagger} H + \lambda (H^{\dagger}H)^2 
  + \frac{a_1}{2} H^{\dagger} H S  
    + \frac{a_2}{2} H^{\dagger} H S^2 + \frac{b_2}{2} S^2 + \frac{b_3}{3} S^{3} + \frac{b_4}{4}S^4 , \quad 
  \label{eq:v}
\end{eqnarray} 
where $S=v_s + s$ is the new singlet scalar and  $H^{\text{T}} = (G^+, (v_{\text{EW}}+ h + i G^0)/\sqrt{2})$ is the SM Higgs doublet with $v_{\text{EW}}=246~\text{GeV}$. All parameters of the above scalar potential are real. Using the tadpole conditions of the potential, we can replace two of these parameters ($\mu$, $b_2$). The physical scalars in the model can be obtained by the rotation
\begin{eqnarray}
h_1 = c_{\theta} h + s_{\theta} s, \quad \quad
h_2 =-s_{\theta} h + c_{\theta} s.  
  \label{}
\end{eqnarray}
We identify $h_1$ as the SM Higgs, $m_{h_1}\sim125$~GeV, and $h_2$ as a heavier scalar resonance. Three more parameters in the scalar potential can be replaced by the masses and
mixing angle of the physical scalars $(m_{h_1}, m_{h_2}, \theta)$. It is usual in the literature to replace $(\lambda, a_1, a_2)$ by the above three physical parameters, with $(b_3, b_4)$ being considered independent parameters. In this paper, we take a different approach and choose the cubic couplings $(a_1, b_3)$ as independent parameters instead. 
We opt for this path since the cubic couplings play the most important role in forming a barrier during the Electroweak Phase Transition (EWPT) and the subsequent production of GW signals. Hence, the unknown free parameters within our setup, which can specify the model completely, are
\begin{eqnarray}
\centering
v_s, \quad \quad m_{h_2}, \quad \quad \theta, \quad \quad a_1, \quad \quad b_3 .
\end{eqnarray}
In the subsequent parameter scan for our analysis, we consider both positive and negative values of cubic couplings ($a_1$,  $b_3$) and the singlet vev $v_s$, but scan only positive values of $\sin \theta$ without any loss of generality. 

One can impose two general categories of constraints on the parameter space of the xSM. The first set of constraints are theoretical. They include the stability of the EW vacuum, boundedness of the potential from below, and perturbative unitarity of $2 \rightarrow 2$ scattering processes. All the other constraints are phenomenological. In particular, Higgs signal strength measurements  constrain the mixing angle $\theta$~\cite{Khachatryan:2016vau,Sirunyan:2018koj}. EW precision measurements, such as corrections to $m_W$~\cite{Lopez-Val:2014jva,Robens:2015gla} and the oblique $S,T,U$ parameters~\cite{Peskin:1991sw,Hagiwara:1994pw},  constrain the model in $(m_{h_2}, \theta)$ plane
at one-loop level. The $W$-mass measurement typically  provides the strongest bounds. For the details of the model and the impact of the various limits on the parameter space, we refer the reader to our previous paper~\cite{Alves:2018jsw}.

 The analysis of the phase transition and eventual calculation of gravitational waves starts with the finite temperature effective potential. This can be obtained in the high-temperature approximation, where gauge-independence is explicitly maintained~\cite{Patel:2011th}. The resulting effective potential takes the same form except that the parameters 
$\mu$ and $b_2$ now become temperature-dependent. Additional contributions to the cubic term are of secondary importance as the tree-level cubic terms are assumed to dominate the barrier. If the effective potential can accommodate a first-order EWPT, a stochastic background of gravitational waves can be generated through the nucleation and collision of the electroweak bubbles in the super-cooled plasma consisting of relativistic particles~\cite{Kosowsky:1991ua,Kosowsky:1992rz,Kosowsky:1992vn,Huber:2008hg,Jinno:2016vai,Jinno:2017fby,Hindmarsh:2013xza,Hindmarsh:2015qta, Guo:2020grp}. The parameters characterizing the dynamics of the phase transition are:
$(T_c$, $T_n$, $\alpha$, $\beta$, $v_w)$, where $T_c$ is
the critical temperature when the would-be true vacuum is 
degenerate with the meta-stable one; $T_n$ is the nucleation
temperature when there is approximately one bubble per 
Hubble volume; $\alpha$ is the energy density released due to the phase transition normalized by the radiation energy density of the universe; $\beta$ is 
roughly the inverse time scale for the phase transition; and
$v_w$ is the bubble wall velocity.

The EWPT would result in gravitational waves naturally falling within the milli-Hertz frequency band  and 
can potentially be detected by future space-based detectors~\cite{Grojean:2006bp}, of which many have been proposed~\cite{amaroseoane2017laser,Crowder:2005nr,Yagi:2011wg,Gong:2014mca,Luo:2015ght}. Different from a chirp signal coming from 
binary black hole 
mergers, the most important feature of the  gravitational wave signal generated from a cosmological first-order phase transition is its stochastic origin.
Therefore, its detection requires at least a pair of independent interferometers, with the strength of the signal represented by the signal-to-noise ratio (SNR)~\cite{Allen:1997ad,Romano:2016dpx,Christensen:2018iqi,Caprini:2015zlo}. 

\subsection{Emergence of Blind Spots}

We now turn to an investigation of blind spots in this setting.  First-order phase transition can be realized for negative cubic and positive quadratic terms, also keeping the potential  bounded from below. It is illuminating to trade the cubic couplings ($a_1, b_3$) for the couplings of $h_2 h_1 h_1$ $(g_{211})$ and $h_1 h_2 h_2$ $(g_{122})$
\begin{eqnarray}
g_{211} & = &  \dfrac{\sin \theta}{2} \bigg[- \dfrac{\sin 2\theta}{2} b_3 + \Big( \dfrac{v_{\text{EW}}}{v_s}(1- 3 \cos^2 \theta) + \dfrac{3}{4} \dfrac{v_{\text{EW}}^2}{v_s^2} \sin 2 \theta \Big) \dfrac{a_1}{2} +
\nonumber \\
&&  (2 m^2_{h_1} + m^2_{h_2}) \Big(\dfrac{1}{2 v_s} \sin 2\theta - \dfrac{\cos^2 \theta}{v_{\text{EW}}}  \Big) \bigg] \, ,
\nonumber \\
g_{122} & = &  \dfrac{\cos \theta}{2} \bigg[- \dfrac{ \sin 2\theta}{2} b_3 + \Big( - \dfrac{v_{\text{EW}}}{v_s}(1- 3 \sin^2 \theta) + \dfrac{3}{4} \dfrac{v_{\text{EW}}^2}{v_s^2} \sin 2 \theta \Big) \dfrac{a_1}{2} +
\nonumber \\
&&  ( m^2_{h_1} + 2 m^2_{h_2}) \Big(\dfrac{1}{2 v_s} \sin 2\theta + \dfrac{\sin^2 \theta}{v_{\text{EW}}}  \Big) \bigg] \, .
\label{Eq:triple_coupling}
\end{eqnarray}
The triple Higgs coupling $g_{111}$ is also  relevant
\begin{eqnarray}
g_{111} & = &  \dfrac{1}{2} \bigg[- \sin^3 \theta \dfrac{b_3}{3} + \sin^2 \theta \Big( -\dfrac{v_{\text{EW}}}{v_s} \cos \theta + \dfrac{1}{2} \dfrac{v_{\text{EW}}^2}{v_s^2} \sin \theta \Big) \dfrac{a_1}{2} +
\nonumber \\
&&   m^2_{h_1}  \Big(\dfrac{ \sin^3 \theta}{v_s} + \dfrac{\cos^3 \theta}{v_{\text{EW}}}  \Big) \bigg] \, .
\label{Eq:triple_coupling2}
\end{eqnarray}

A blind spot for resonant di-Higgs production is obtained when  $g_{211}$ is depleted, $g_{211}\approx 0$, while a different coupling can still provide a barrier for a first-order phase transition with appreciable gravitational wave signals. When we fix $\sin \theta$ and $v_s$, we get a linear combination of $a_1$ and $b_3$ on the right hand side of Eq.~\eqref{Eq:triple_coupling}. One can readily evaluate the limiting case $g_{211}\rightarrow 0$  with
\begin{eqnarray}
m_{h_2} & = & \dfrac{1}{2} \bigg( - 8 m_{h_1}^2 + 4 b_3 v_s - \dfrac{3 a_1 v_{\text{EW}}^2}{v_s} + \dfrac{ 4 b_3 v_s^2 + a_1 v_{\text{EW}}^2 (1- 2 \tan^2 \theta)}{v_{\text{EW}} \tan \theta -v_s}\bigg)^{1/2}\,.
\label{Eq:mh2_blindspot}
\end{eqnarray} 
Although the $g_{211}\rightarrow 0$ regime is allowed by the constraints on the model, there is no condition favoring this parameter space point. The conditions for obtaining large SNR for GWs observation will, however, alter this scenario as  we promptly discuss.

\begin{figure}[t!]
    \centering
    \includegraphics[width=0.45\textwidth]{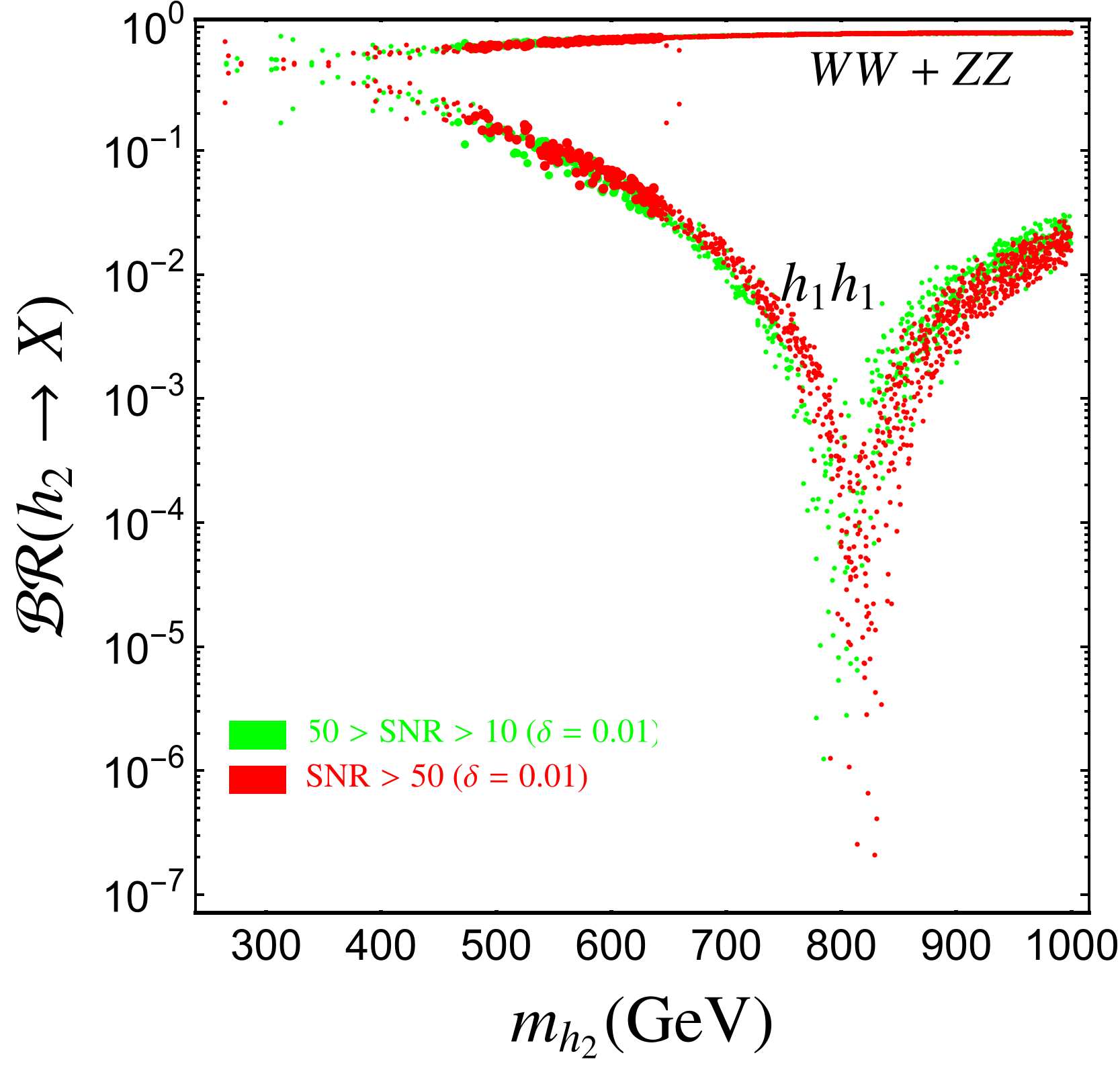}
    \includegraphics[width=0.45\textwidth]{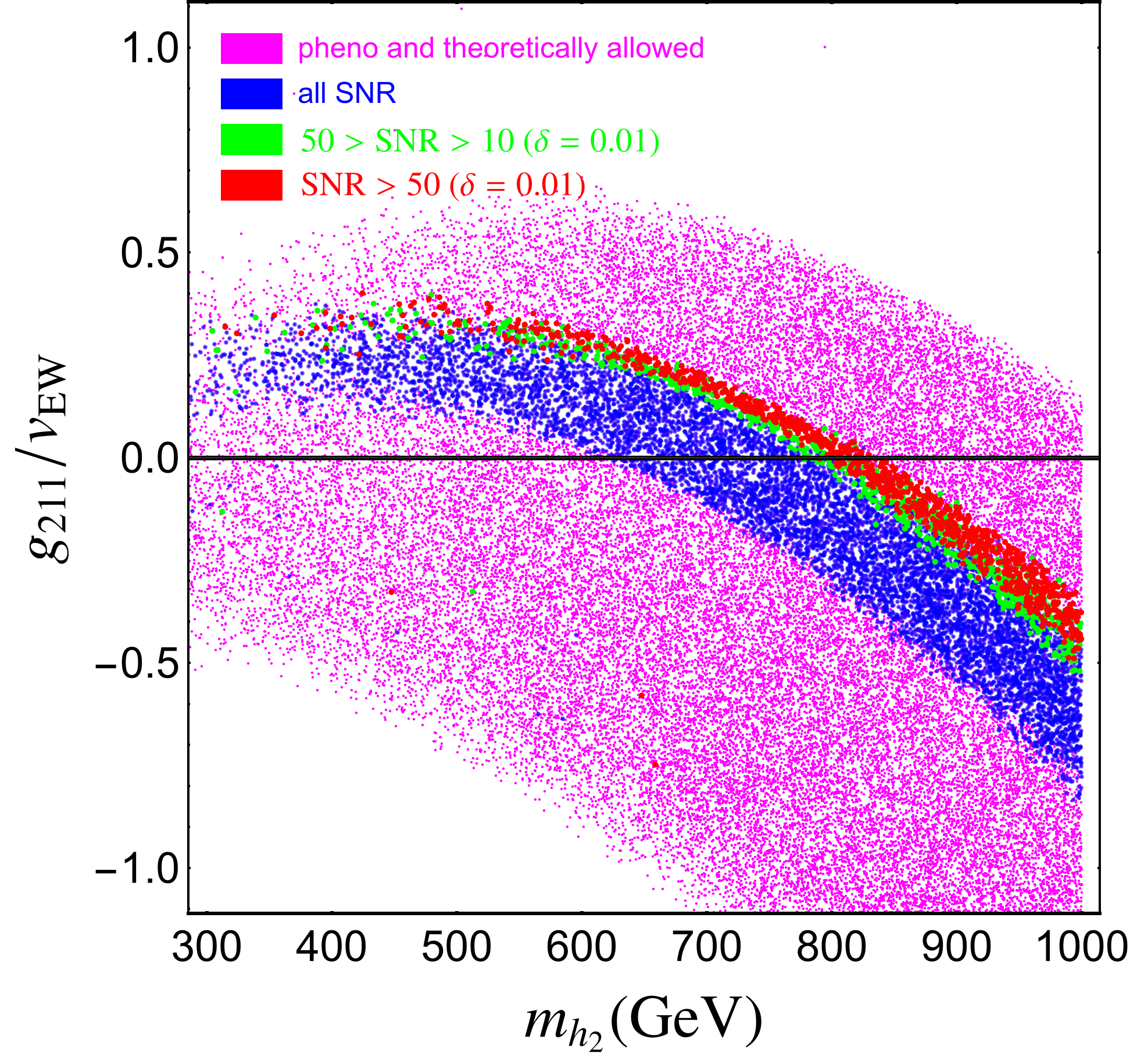}
    \caption{Left panel: Branching ratios of the heavier scalar $h_2$ that can accommodate  first-order phase transition with signal-to-noise ratio $10<\text{SNR}<50$ (green), and $\text{SNR}>50$ (red).
    Right panel: The coupling $h_2 h_1 h_1$ for phenomenologically and theoretically allowed points (pink), and points that satisfy first-oder phase transition with all SNR (blue), $10<\text{SNR}<50$ (green), and $\text{SNR}>50$ (red). We assume $\sin\theta=0.2$.}
    \label{fig:br}
\end{figure}

In Fig.~\ref{fig:br} (left panel),  we show the branching ratios of the new scalar  to SM Higgs pair ${h_2 \rightarrow h_1h_1}$ and to di-boson pair ${h_2 \rightarrow VV}$, where $V=Z,W$, assuming $\sin \theta = 0.2$. The parameters $(v_s,a_1,b_3)$ are  allowed to vary.
The points are compatible with all phenomenological and theoretical consistency requirements for the model (we refer to Ref.~\cite{Alves:2019igs}  for an exhaustive discussion of  phenomenological  constraints). The stochastic GW signals at LISA are denoted in two different SNR regions:
 $10< \text{SNR}  < 50$ (green) and  $\text{SNR} > 50$ (red). We have applied a reduction factor $\delta = 0.01$ for all  points in calculating the SNR to be conservative~\cite{Alves:2019igs}, considering the recently observed reduction in  gravitational wave production from sound waves~\cite{Cutting:2019zws}.
It is evident that the di-Higgs branching ratio falls precipitously near $m_{h_2} \sim 800$ GeV, even though a large stochastic gravitational wave signal is obtained. 
Since the  allowed parameter space displays suppressed branching ratios over a wide range of  $m_{h2}$, going further beyond the limiting point $m_{h2}\sim 800$~GeV, we can foresee challenging collider prospects for this channel, resulting in a significantly large di-Higgs blind spot regime. The LHC prospects will be derived in Sec.~\ref{sec:collider}.

In Fig.~\ref{fig:br} (right panel), we directly plot $g_{211}$. The coupling is normalized by $v_{\text{EW}}$. The pink regions denote parameter space points compatible with all theoretical and phenomenological requirements. The blue points denote models that give rise to a first-order phase transition and gravitational waves of \textit{any} strength. The horizontal line indicates the vanishing coupling $g_{211} \rightarrow 0$. It is clear that while phenomenologically acceptable pink points are distributed with relative uniformity, exploring the extra degrees of freedom in the Higgs potential shown in Eq.~\ref{Eq:triple_coupling}, the requirement of a successful first-order phase transition (i.e.,
with the condition for defining $T_n$ satisfied) already restricts the parameter space to a narrow region in $g_{211}-m_{h_2}$. This space becomes even more constricted as the requisite SNR becomes larger. 

Before analyzing the reasons for this parameter constriction in more detail, a few comments about the outlier points  with large but negative $g_{211}/v_{\text{EW}}$, on the right panel of Fig.~\ref{fig:br}, are in order. Interestingly, these few points possess large negative $v_s$, while the band structure is formed by positive $v_s$ points. Although we start our scan with equal number of points with positive and negative values of $v_s$, phenomenological and theoretical constraints overwhelmingly prefer positive $v_s$ points. We obtain only $\sim 5 \%$ points with negative $v_s$. Successful completion of first-order EWPT further disfavors negative $v_s$ points, and they are only $\sim 0.25 \%$ of total number of points that undergo first-order EWPT. Imposing the condition of requiring strong SNR does not change this ratio significantly. 
Obviously the outliers require positive $a_1$ and $b_3$ to form a barrier since they have $v_s < 0$. {We find} that $a_1$ and $b_3$ enjoy almost a linear relationship for the outliers. In contrast, the points in the bulk prefer negative $a_1$ but are uniformly distributed in $b_3$.   
The relationship between $a_1$ and $b_3$ plays an important role in forming the blind spot, as we will discuss in the subsequent text. Also, we found that a modest $\sim1\%$ of points undergo two-step phase transition~\cite{Alves:2018jsw}. The mechanism of two-step phase transition is different from one-step transition, thus the parameter space preferred by those points will be naturally different. Because of their suppressed likelihood, different physics origin and spectral shapes, we do not further discuss those points in this paper. Finally, we further point out a caveat on our calculation for the nucleation criteria at a temperature very close to the minimum of the action. These points naturally have a small first derivative $\beta$ and thus large SNR. While a better treatment might be obtained by using the second derivative $\beta_2$ (see, e.g., \cite{Hindmarsh:2019phv,Guo:2020grp}),  the true observable that determines the spectral shape is the mean bubble separation, whose relation with $\beta$ and $\beta_2$ needs to be studied case by case as analyzed in Ref.~\cite{Guo:2020grp}.
We leave such a detailed analysis for a future study.

\begin{figure}[!t]
    \centering
    \includegraphics[width=0.45\textwidth]{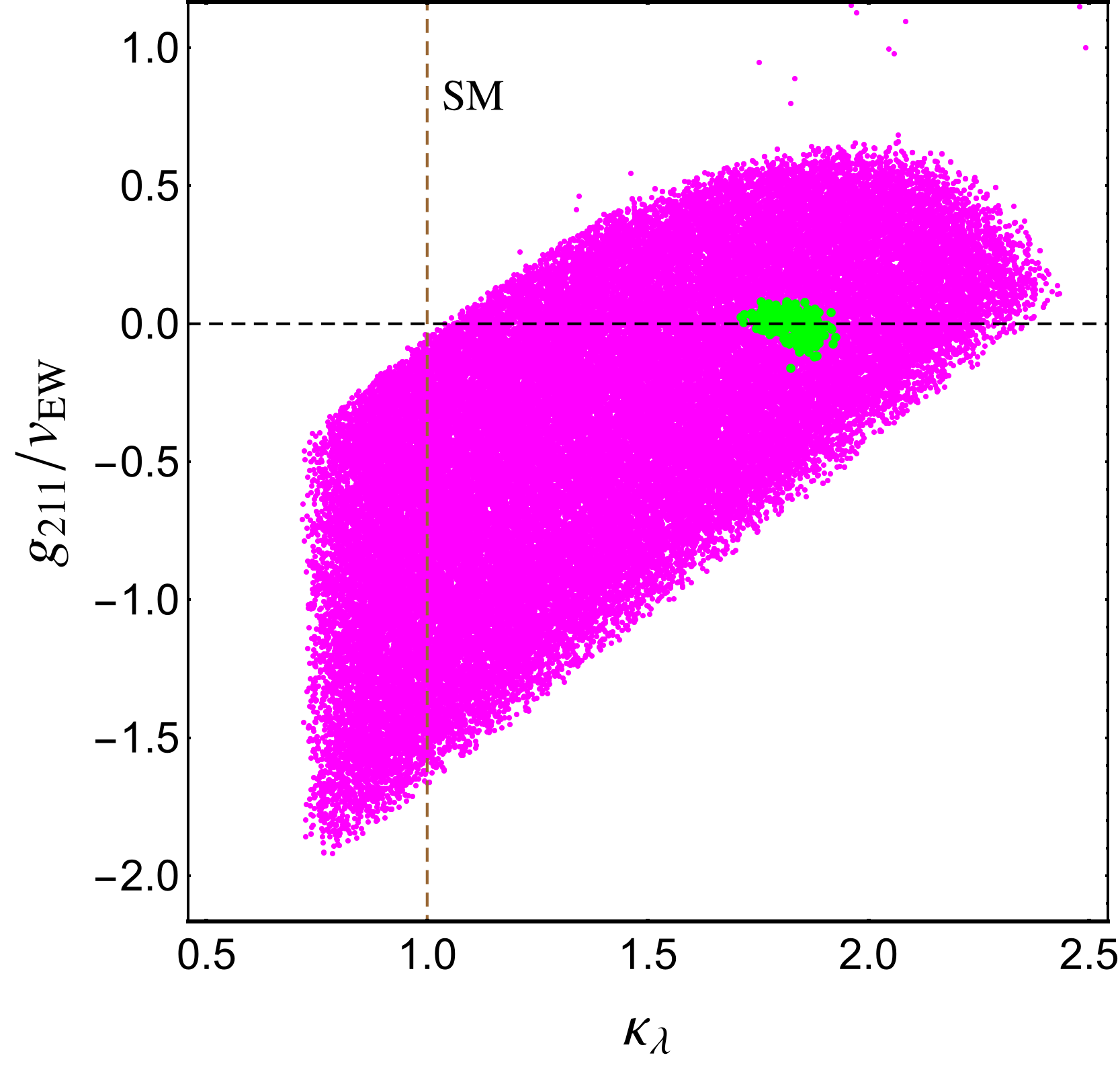}
    \includegraphics[width=0.45\textwidth]{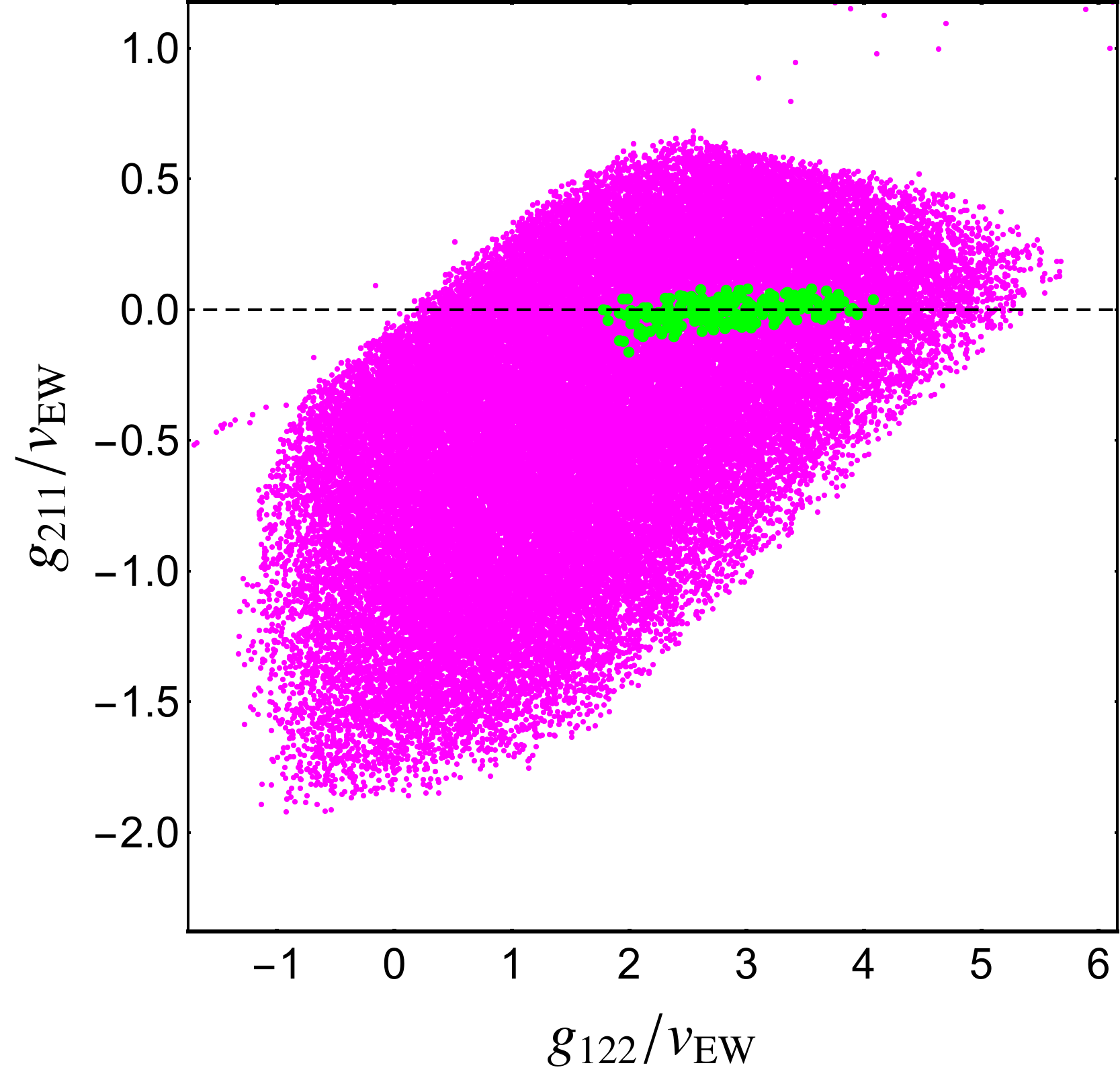}
     \\
     \includegraphics[width=0.45\textwidth]{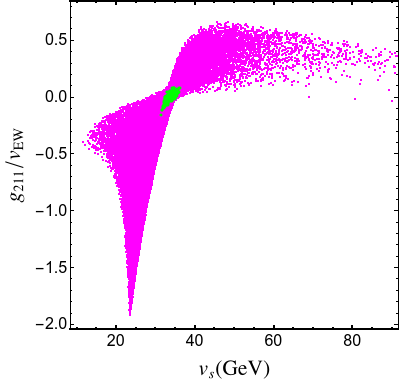} 
     \includegraphics[width=0.45\textwidth]{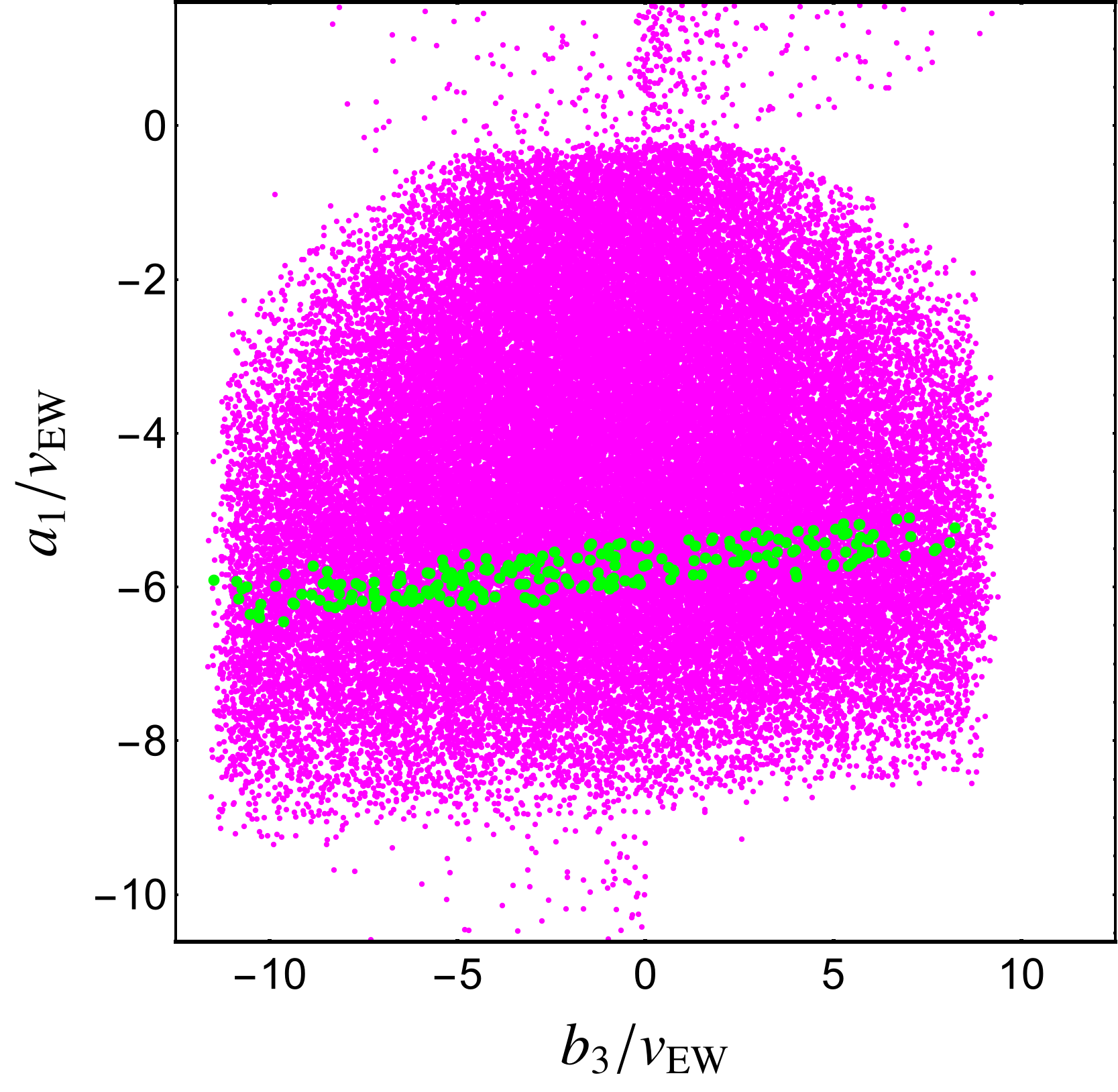} 
    \caption{Points in $(g_{211}/v_{\text{EW}}, \kappa_{\lambda})$ (top-left), $(g_{211}/v_{\text{EW}},g_{122}/v_{\text{EW}})$ (top-right), $(g_{211}/v_{\text{EW}},v_s)$ (bottom-left) and $(a_1/v_{\text{EW}}, b_3/v_{\text{EW}})$ (bottom-right) planes. The green points satisfy $780~\text{GeV}<m_{h_2}<840$ GeV and $\text{SNR}>10$, assuming a universal suppression factor of $0.01$. These are part of the green and red points near the 
    blind spot regime depicted in Fig.~\ref{fig:br}.
    The pink regions correspond to the  pink points of Fig.~\ref{fig:br}, i.e., all points compatible with phenomenological and theoretical constraints.
    }
    \label{fig:couplings}
\end{figure}

 Let us now focus primarily on the points that have suppressed $h_2 \rightarrow h_1 h_1$ branching ratio. In fact, for the red points, i.e., those with SNR greater than $50$, the fraction that has ${\mathcal{BR}(h_2 \rightarrow h_1 h_1) <10^{-2}}$ $(10^{-3})$ is about $56\%$ $(16\%)$. Restricting to the  mass window $780~\text{GeV}<  m_{h_2} < 840$~GeV, we obtain 100\%  (91\%) of the simulated events with $\mathcal{BR}(h_2 \rightarrow h_1 h_1) <$ $10^{-2}$ $(10^{-3})$. Whereas the di-Higgs branching ratio gets further suppressed within the mass window around $g_{211} \rightarrow 0$, a wider span of $m_{h_2}$ will remain beyond the ambit of resonant di-Higgs searches at the LHC. The reason for this phenomenological effect is manifest in the right panel of Fig.~\ref{fig:br}, where the red band crosses the $g_{211}=0$ line with a small slope, and hence, $|g_{211}|$ does not attain large values within the GW motivated parameter space.

 In  Fig.~\ref{fig:couplings} (top-left panel), we show the blind spot in the space of couplings ($g_{211}/v_{\text{EW}}$, $\kappa_{\lambda}$) where $\kappa_{\lambda} \equiv g_{111}/g_{111}^{\text{SM}}$. The mass of $h_2$ has been restricted to the limiting range $780~\text{GeV} < m_{h_2} < 840$~GeV. 
 The pink regions correspond to the pink region of the right panel of Fig.~\ref{fig:br}.  The green dots correspond to  points with SNR greater than $10$, assuming a universal suppression factor of $0.01$. Remarkably,  these conditions imply in a narrow range for the  Higgs self-coupling $1.7\lesssim\kappa_\lambda\lesssim 1.9$. It is well known that the non-resonant di-Higgs production cross-section becomes smaller for $\kappa_{\lambda} > 1$ due to the increasingly destructive interference of the triangle and box diagrams with the minimum being at $\kappa_{\lambda}\approx 2.4$~\cite{Goncalves:2018qas}. This implies that  the non-resonant di-Higgs cross-section is also suppressed to almost half of the SM rate, making it unlikely to be probed by non-resonant searches either. We also show the other triple coupling in the $(g_{211}/v_{\text{EW}}, g_{122}/v_{\text{EW}})$ plane in the top right panel of the same figure. We see that a rather large range of $1.8 \lesssim g_{122}/v_{\text{EW}} \lesssim 4.1$ is available to provide a first-order phase transition in the limit $g_{211} \rightarrow 0$.

\begin{figure}[t]
    \centering
    \includegraphics[width=0.3\textwidth]{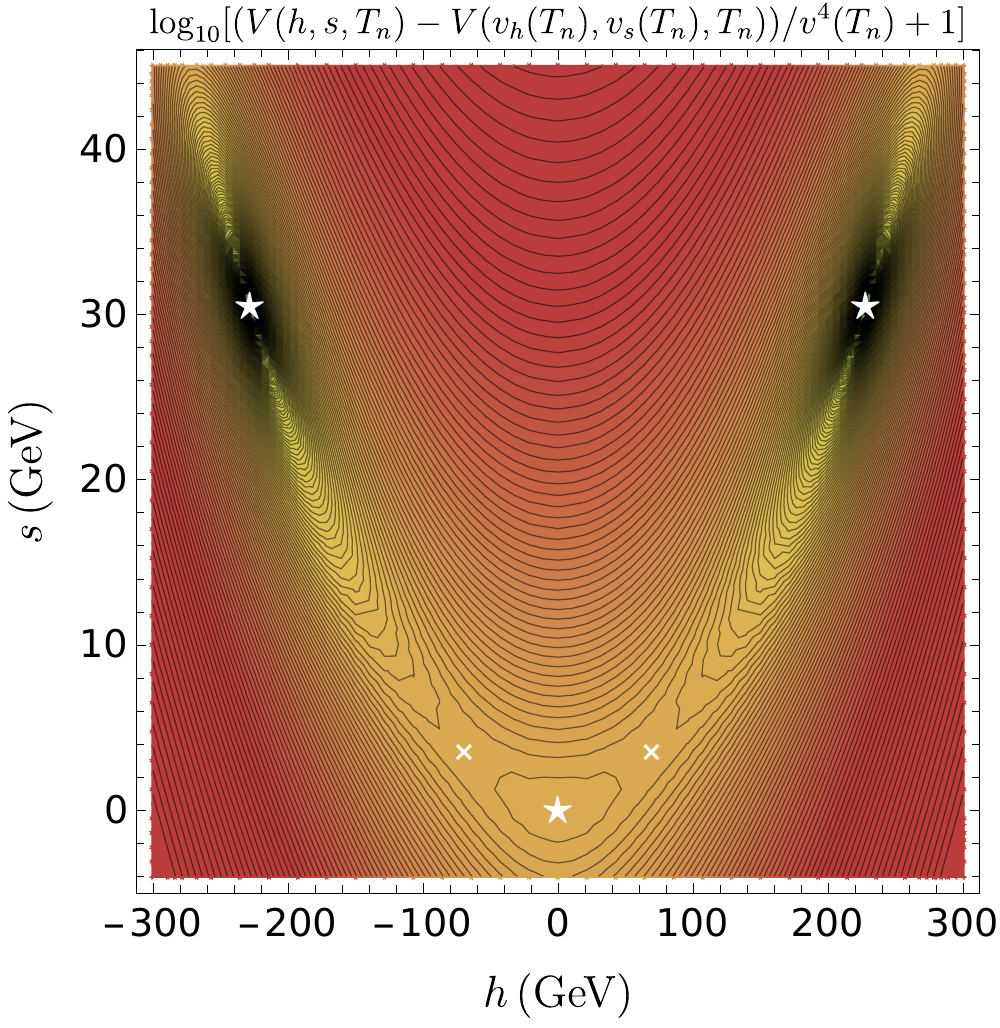}
     \includegraphics[width=0.3\textwidth]{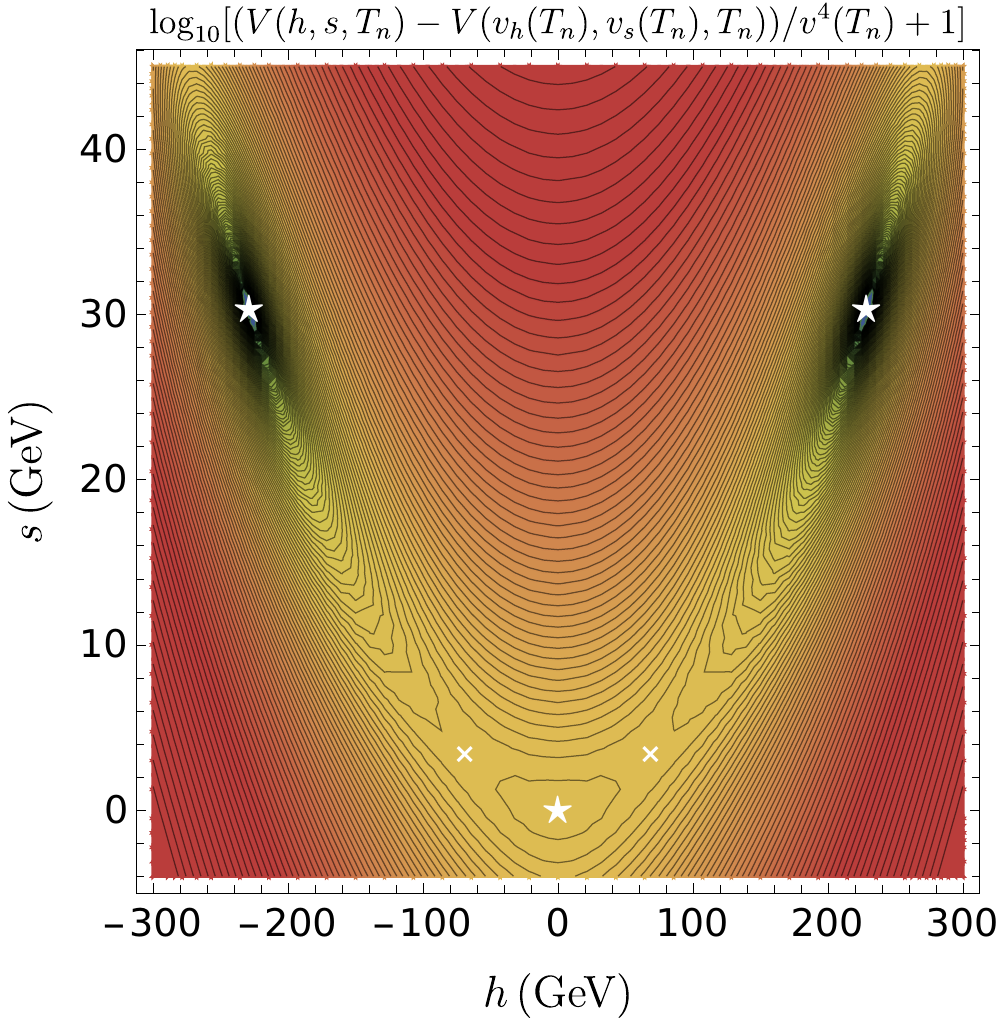}
        \includegraphics[width=0.36\textwidth]{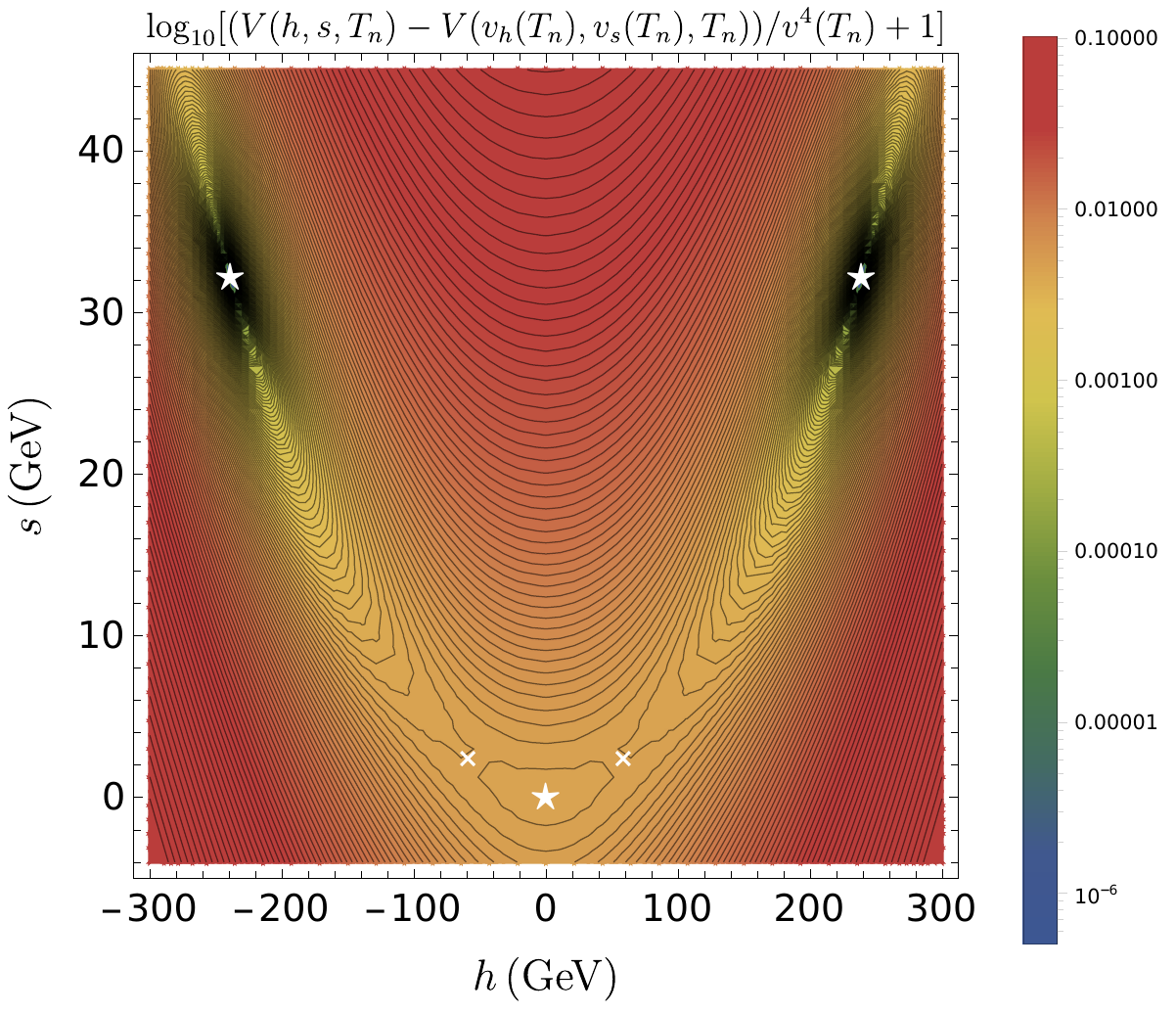}
    \caption{The shape of the potential at respective nucleation temperatures for three benchmark points with $\text{SNR}> 10$. The potentials are shown in the logarithmic scale in the $z-$axis and normalized as $\log_{10}\big[ (V(h,s,T_n) -  V(v_{EW}(T_n), v_s(T_n),T_n)/{v(T_n)^4} \, + \, 1\big]$, where $v(T_n)\equiv (v_h^2(T_n) + v_s^2(T_n))^{1/2}$. The input parameters $(v_s, m_{h_2}, \sin \theta, a_1, b_3 )$ for the three points are $( 34.1\, \, \text{GeV}, 818 \, \text{GeV}, \, 0.2, \,-1394 \, \text{GeV}, \, 612 \, \text{GeV})$, $( 33.8 \,\text{GeV},\, 814 \, \text{GeV},\, 0.2,\,-1408 \, \text{GeV},\, -9.56 \, \text{GeV})$ and $( 33.6 \,\text{GeV},\, 813 \, \text{GeV},\, 0.2,\,-1468 \, \text{GeV},\, \\-821 \, \text{GeV})$, respectively. They are very similar except for $b_3$. The corresponding nucleation temperatures are $T_n=36.9$~GeV, 35.8 GeV and 23.2 GeV. The minima (maxima) of the potential are highlighted by white stars (crosses). Although the $b_3$ values are significantly different for these points, the shape of the potential for them are nearly identical {around the relevant region for phase transition.}
    }
    \label{fig:aliens}
\end{figure}

We revisit Eqs.~\eqref{Eq:triple_coupling}-\eqref{Eq:mh2_blindspot} to understand why imposing the condition of detectable SNR  results in a predictive value of  $g_{111}$ at the blind spot. As we already mentioned, for a fixed $\sin \theta$ and $v_s$, and hence for a particular linear combination of $a_1$ and $b_3$, we can predict  $m_{h_2}$ from Eq.~\eqref{Eq:mh2_blindspot} where $g_{211}=0$. Similarly, for this specific combination of $a_1$ and $b_3$, we  can  predict a value of $g_{111}$ using Eq.~\eqref{Eq:triple_coupling2}, provided $\sin \theta$ and $v_s$ remain fixed. For our scan results shown in Fig.~\ref{fig:couplings}, although $\sin \theta$ is fixed at 0.2, $v_s$ is not. So, a large variation in $v_s$ will clearly not be predictive for $g_{111}$. However, we have already shown in our previous study~\cite{Alves:2018jsw} that a detectable GW signal favors $v_s$ in a narrow range of $20 - 50$~GeV for all phenomenologically allowed $\sin \theta$. For $\sin \theta =0.2$, the allowed region is even narrower ($27 - 45$ GeV). It is  evident from the distribution of green dots in Fig.~\ref{fig:couplings}, where $780$ GeV $< m_{h_2} < 840$~GeV,  that $v_s$ is preferred to be almost constant ($32-37$~GeV), resulting in a predictive value of $\kappa_{\lambda}$.  

We alluded before that the bulk points show no preference in $b_3$. In the bottom right panel of Fig.~\ref{fig:couplings},
 we  present the distribution of $a_1$ and $b_3$, normalized by $v_{\text{EW}}$. Clearly, $a_1$ is preferred in a significantly smaller range,
$a_1/v_{\text{EW}}\sim [-6.5,-4.9]$,
when compared to $b_3$, which spans the whole scan range. 
So, we can infer that $b_3$ plays a minimal role in EWPT for the blind spot points. To stress this property, we show in Fig.~\ref{fig:aliens} the shape of the potential at respective nucleation temperatures for three benchmark points with $\text{SNR}> 10$. These three points have similar values of input parameters $(v_s, m_{h_2}, \theta, a_1 )$ and $(T_c,T_n)$ but widely different values of $b_3$. Undoubtedly, the shape of the potential for these points are almost identical around the relevant region for phase transition. 

Obtaining an analytical expression for why the large SNR requirement imposes regimes with constricted parameter regions is not straightforward. The main challenge is obtaining the bounce solution for the  phase transition (see, e.g., Refs.~\cite{Dunne:2005rt,Andreassen:2016cvx} for 
detailed calculations or~\cite{Weinberg:1996kr} for an introduction). Denoting  the fields collectively as $\vec{\phi}$, the bounce solution minimizes the 3-dimensional Euclidean action
\begin{eqnarray}
S_3(\vec{\phi},T) = 4\pi \int r^2 dr \left[ \frac{1}{2} \left(\frac{d \vec{\phi}(r) }{d r}\right)^2 + V(\vec{\phi},T) \right], 
\end{eqnarray}
with the following boundary conditions
\begin{eqnarray}
  \frac{d \vec{\phi}(r)}{d r}\Big\vert_{r=0} = 0, \quad \quad
  \vec{\phi}(r=\infty) =  \vec{\phi}_{\text{outside}} , 
\end{eqnarray}
where $\vec{\phi}_{\text{outside}}$ denotes the vacuum outside the bubble. Analytic solutions to this minimization problem cannot be obtained except for very special potentials. Given the difficulty in obtaining an analytic condition, we now turn to a detailed numerical study of the scalar potential. 

\subsection{Gravitational Wave Production}

\begin{figure}[t!]
    \centering
        \includegraphics[width=0.32\textwidth]{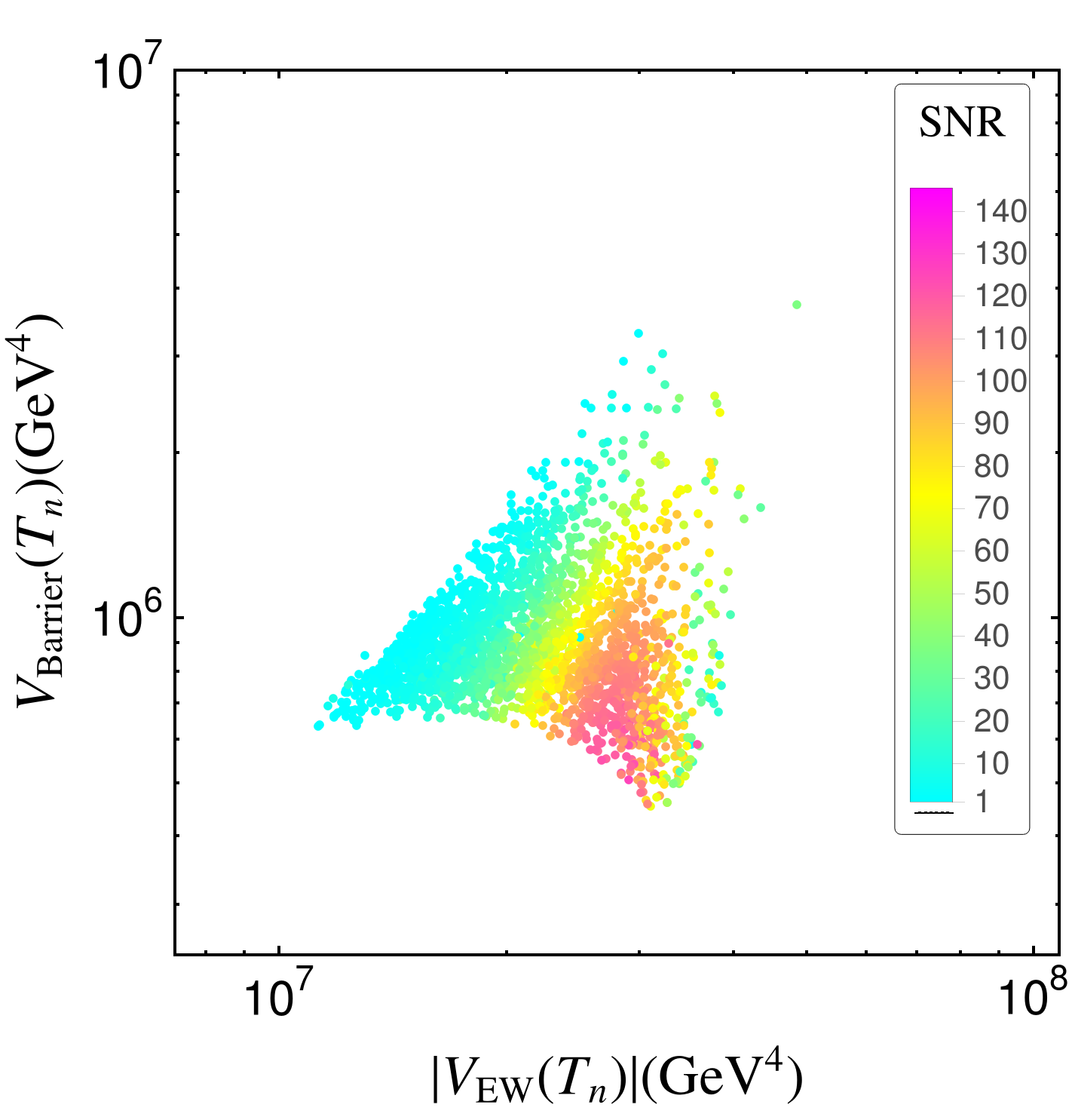}
     \includegraphics[width=0.32\textwidth]{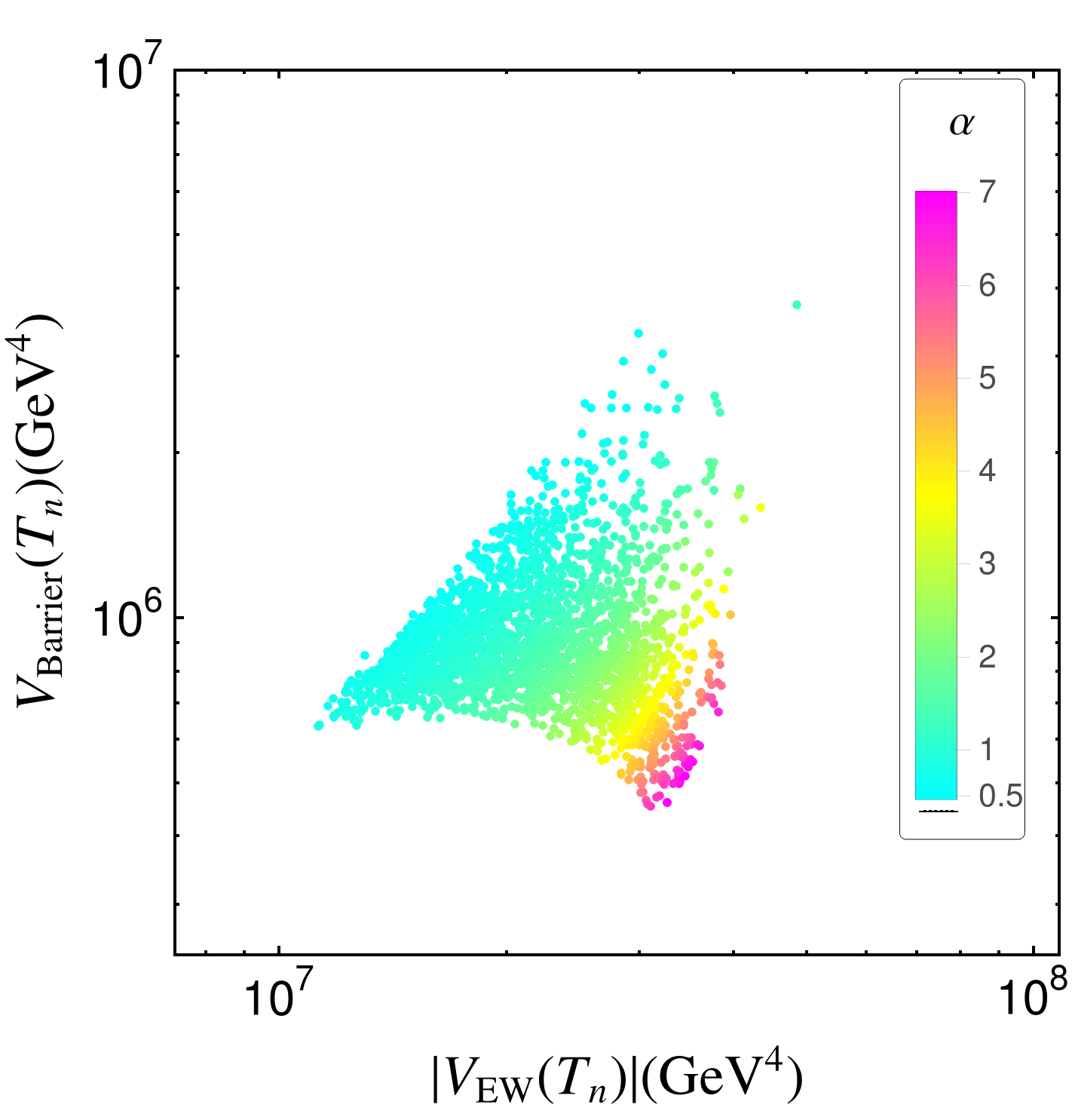}
          \includegraphics[width=0.32\textwidth]{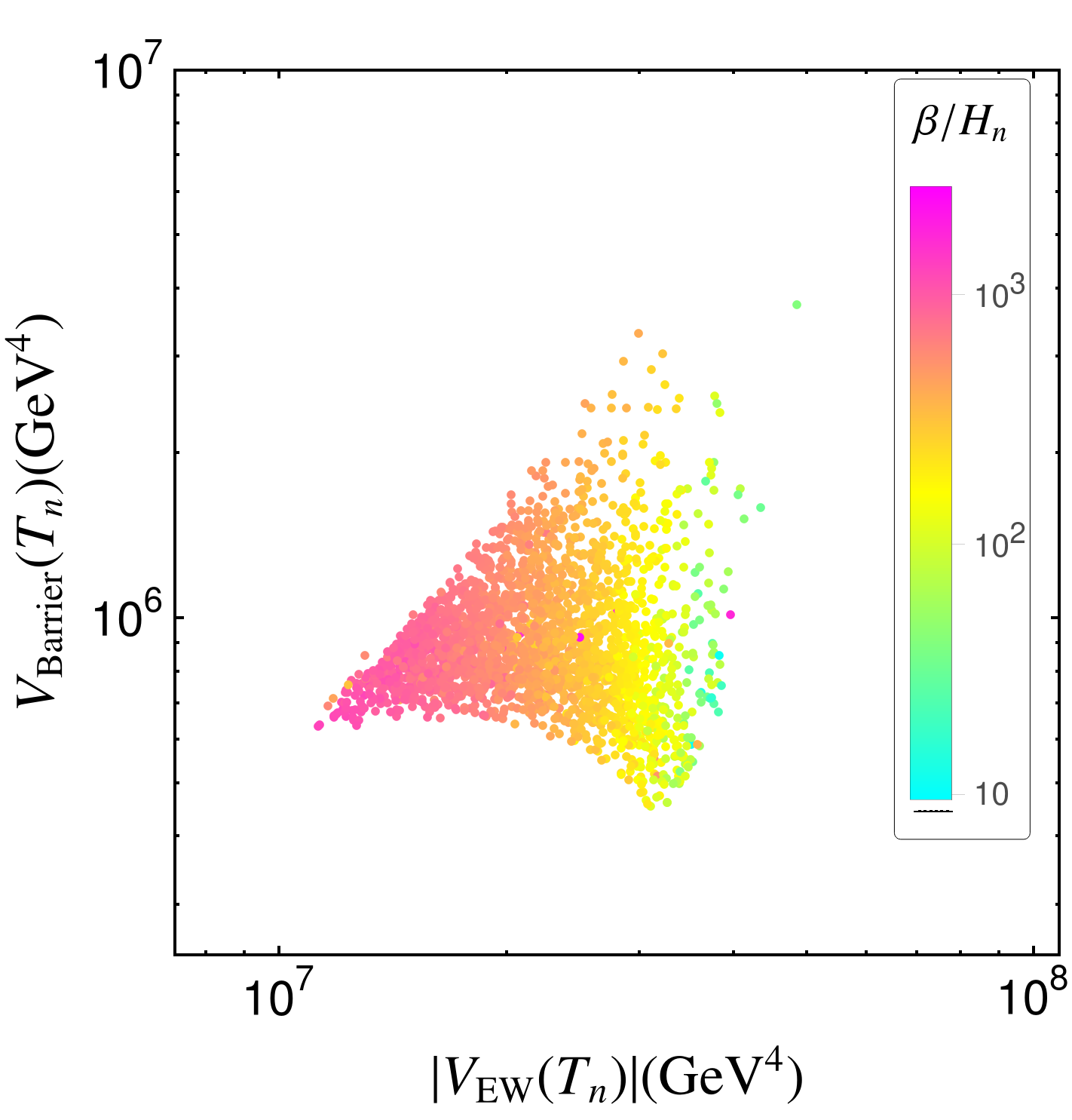}
     \\
     \includegraphics[width=0.32\textwidth]{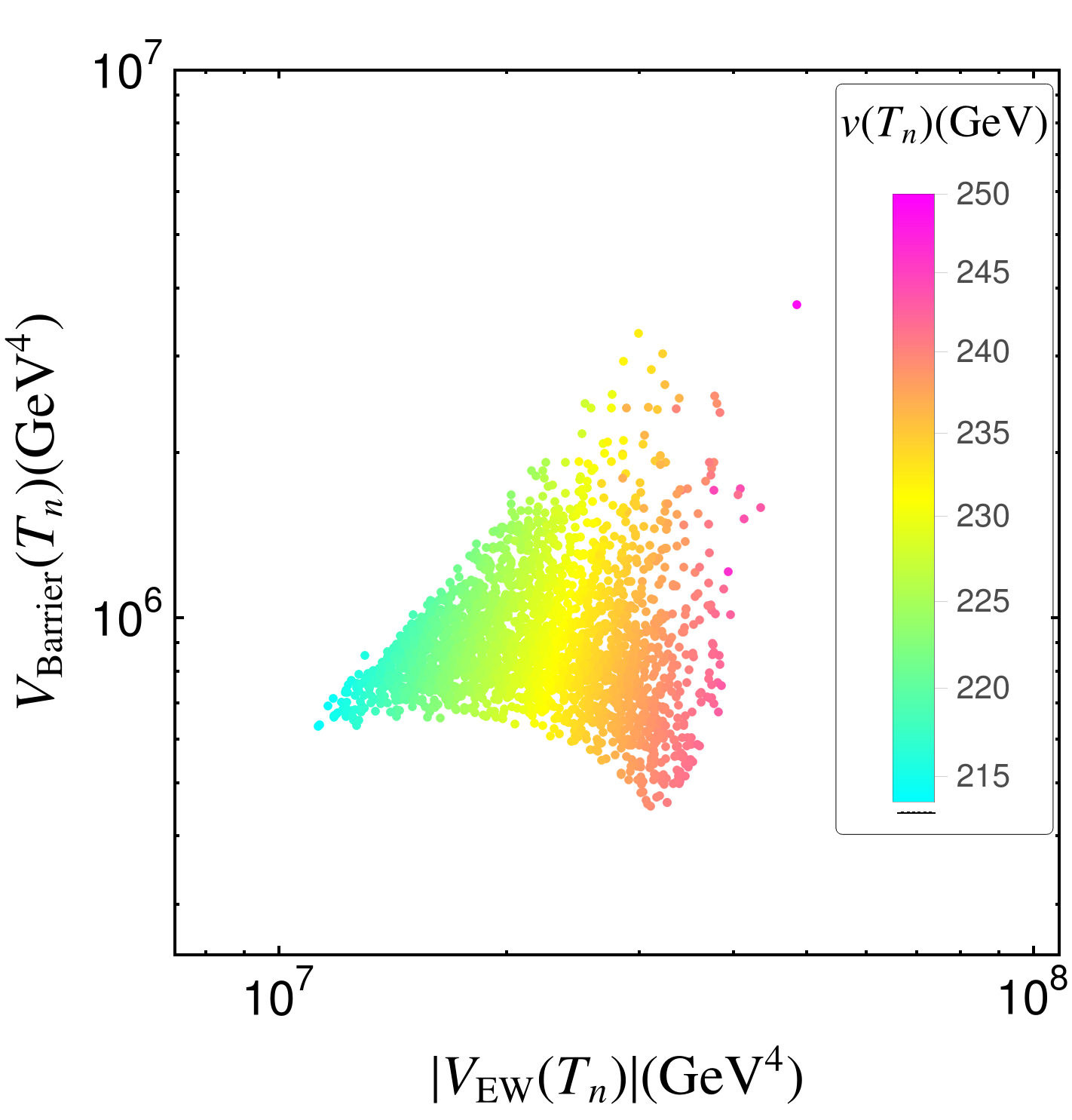}
             \includegraphics[width=0.32\textwidth]{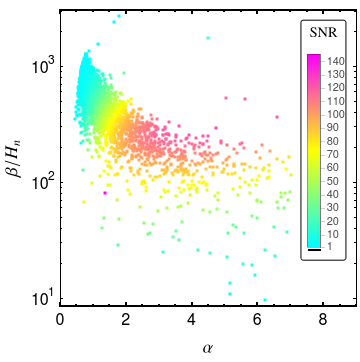}
             \includegraphics[width=0.32\textwidth]{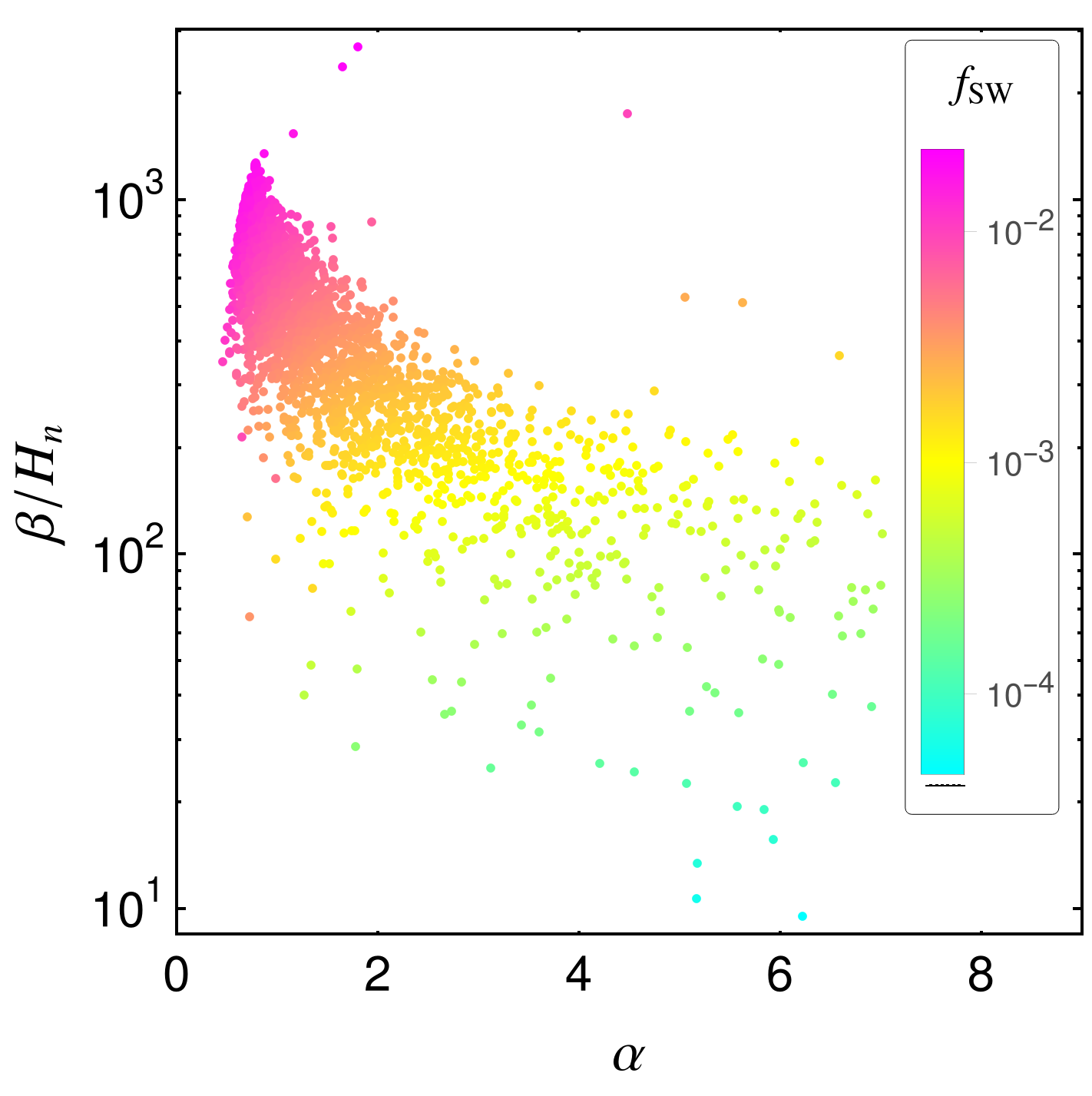}
     \caption{The shape of the potential as characterized by 
     $|V_{\text{EW}}(T_n)|$, $V_{\text{Barrier}}(T_n)$, also
     $\alpha$ and $\beta/H_n$ for the points with SNR larger than 1 for $\sin\theta=0.2$. The bottom right figure
     shows the peak frequency $f_{\text{SW}}$ for the dominant spectrum from the sound waves.
     }
    \label{fig:shape}
\end{figure}

\begin{figure}[b!]
    \centering
     \includegraphics[width=0.42\textwidth]{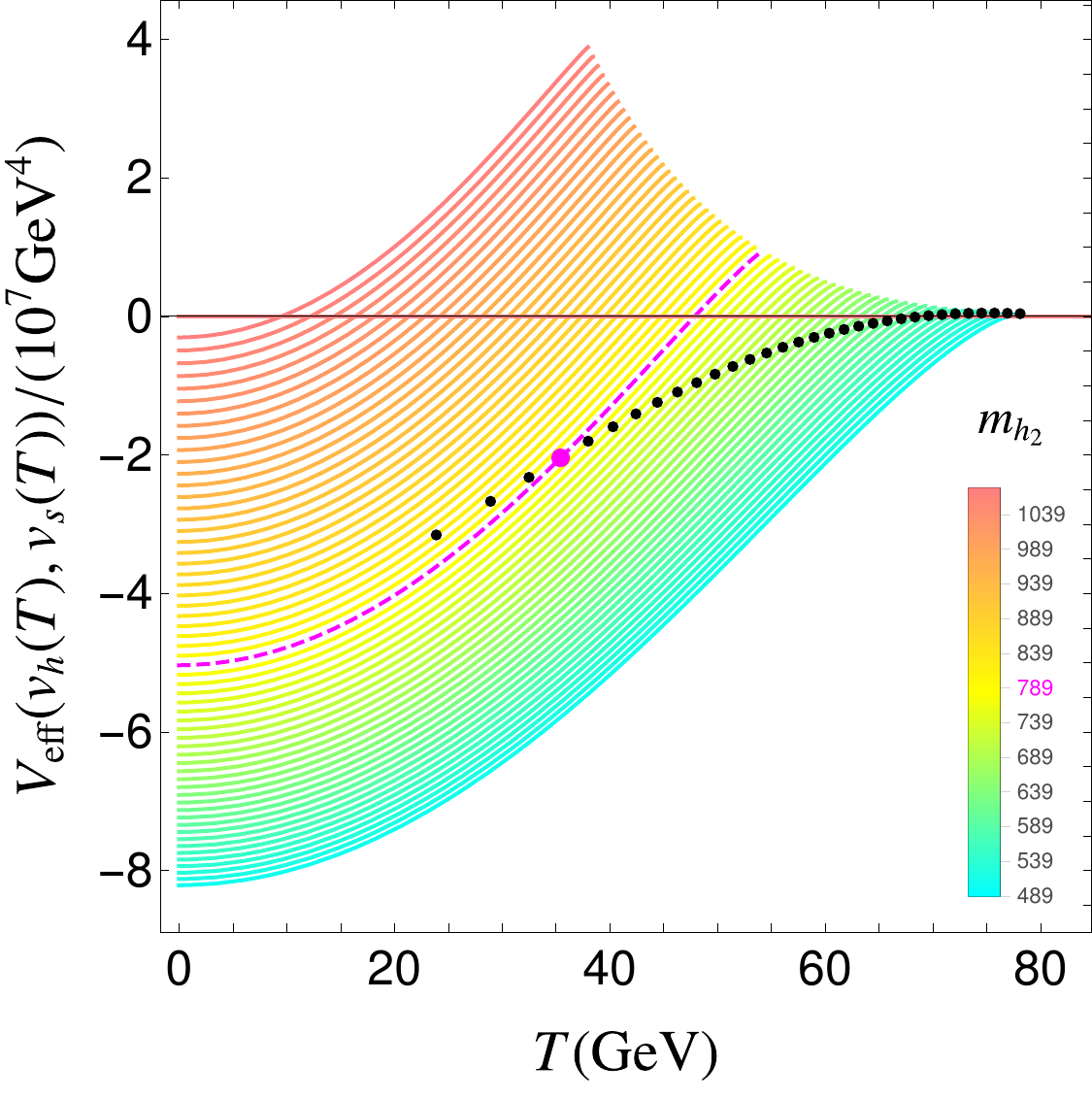}
     \includegraphics[width=0.44\textwidth]{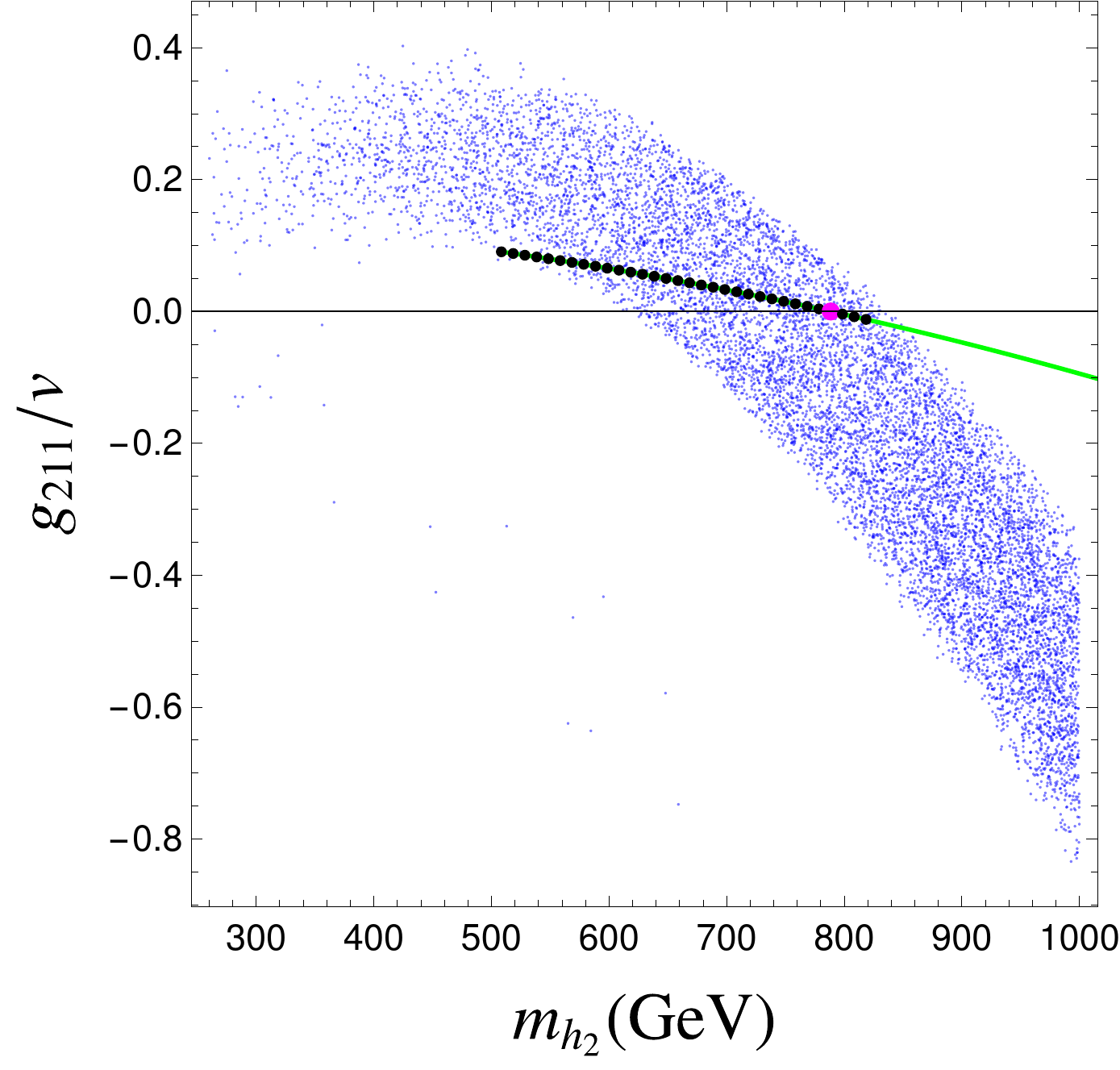}
     \\
      \includegraphics[width=0.405\textwidth]{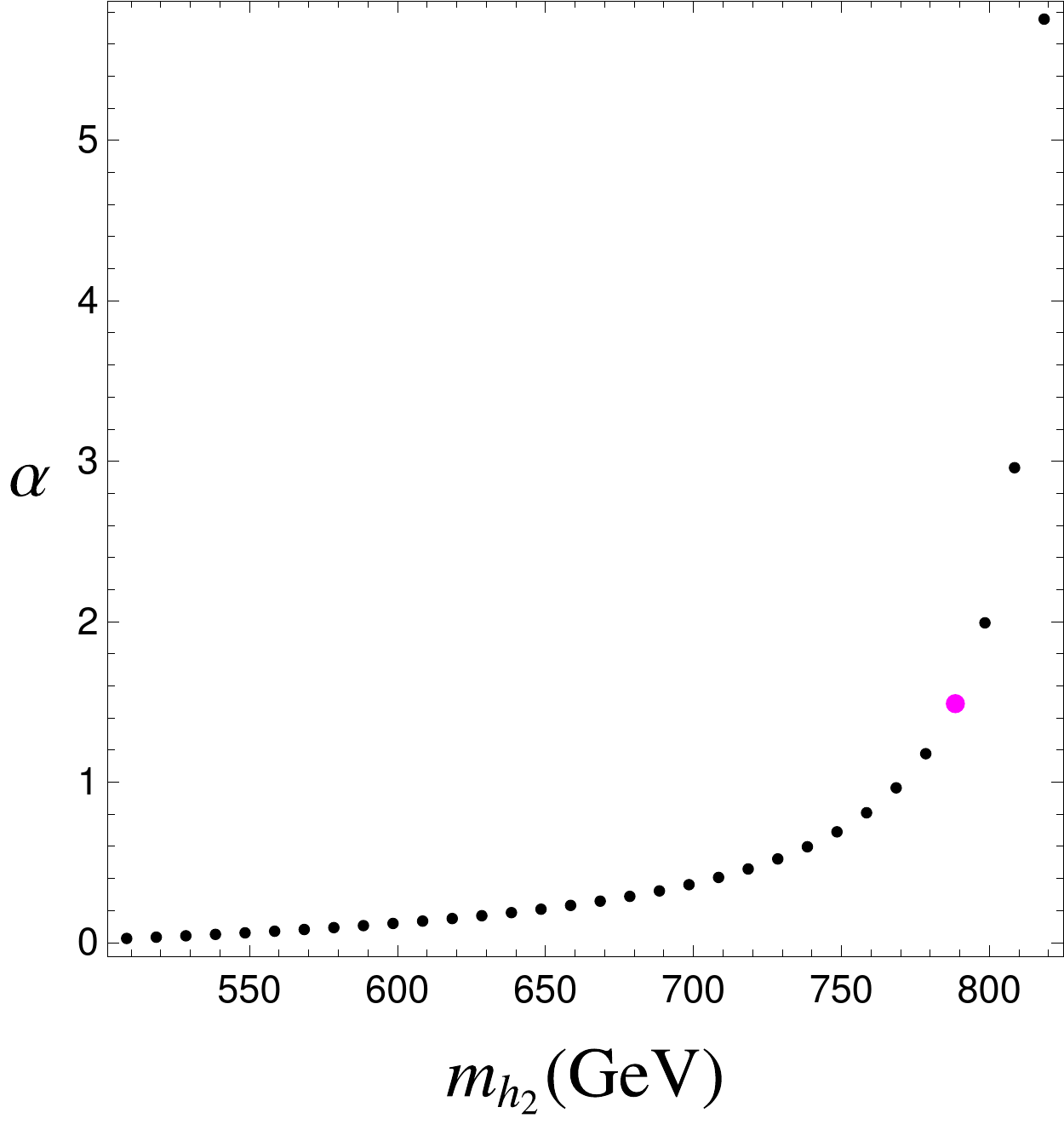}
      \includegraphics[width=0.435\textwidth]{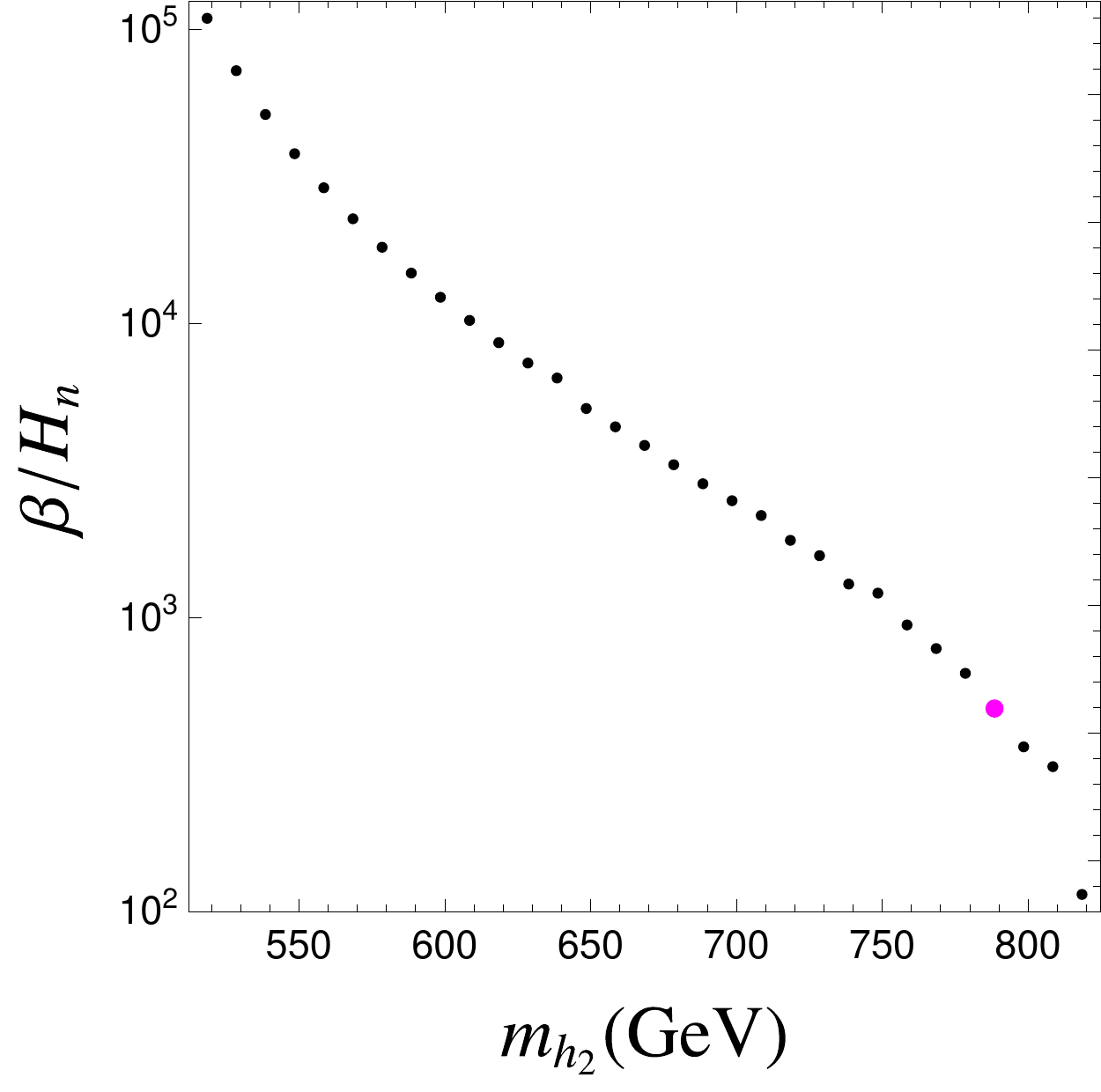}
     \caption{ The behavior of the potential in the vicinity of the blind spot. A benchmark point is chosen, with $\sin\theta=0.2$, $m_{h_2} = 788.5$~GeV, $v_s=32.36$~GeV, $b_3=-2826.11$~GeV, and $b_4=4.19$, which has very small branching to di-Higgs ($g_{211} \sim 0$), and we vary $m_{h_2}$ around this blind point.  The initial benchmark point is shown in magenta in all the panels. As $m_{h_2}$ is varied, the other points in the variation are shown in black if a valid nucleation temperature  $T_n$ can be found and in green if not. In the top left panel, the different curves correspond to the phase histories for all the parameters with the mass shown through a color map. The top right panel shows the coupling $g_{211}$, with the blue band corresponding to the blue points on the right panel of Fig.~\ref{fig:br}. The bottom panels show $\alpha$ and $\beta/H_n$ for the magenta and black points.
     }
    \label{fig:animation}
\end{figure}

In this section, our goal is to study the shape of the potential in the vicinity of a blind spot regime, paying particular attention to the parameters that control the production of gravitational waves. In the top panel of Fig.~\ref{fig:shape}, we show the correlation between the barrier height $V_{\text{Barrier}}(T_n)$ and  the depth of the electroweak vacuum $|V_{\text{EW}}(T_n)|$, where the height and depth are defined with respect to the meta-stable vacuum, for points with SNR larger than 1.
The results are shown for different values of SNR, $\alpha$, $\beta/H_n$, and ${v(T_n)=\sqrt{v_h^2(T_n) + v_s^2(T_n)}}$. We also
show $\alpha$ versus $\beta/H_n$, color-coding SNR.
We use \texttt{CosmoTransitions} to trace the evolution of the phases and obtain  the bounce solutions~\cite{Wainwright:2011kj}.
From these plots, we can see that shapes of the potential 
show the desired behavior for accommodating large SNR for gravitational
waves, i.e., a deeper true vacuum, lower barrier, larger $\alpha$, and smaller $\beta/H_n$~\cite{Weir:2017wfa}. For the plot of $\alpha - \beta/H_n$, with SNR color-coding, there are
points with large $\alpha$ and small $\beta/H_n$ but relatively small SNR,
which might seem counter intuitive as they should, in principle, give a large signal.
The reason is that a very small $\beta/H_n$ would lead to
a correspondingly very small frequency, which then shifts the spectrum 
out of the most sensitive band of LISA and thus results in a smaller SNR~\cite{amaroseoane2017laser}.
The peak frequency $f_{\text{SW}}$ for the dominant source, the sound waves, is shown in a similar $(\alpha - \beta/H_n)$ plot here and serves to explain the behavior of these points.

To further understand the origin of the band structure in 
Fig.~\ref{fig:br} and thus the appearance of the blind spot, 
we choose a benchmark  point  with $m_{h_2} = 788.5 \, \text{GeV}$ 
and vary $m_{h_2}$. 
For each new point obtained by this variation, the resulting phase transition is calculated and presented in Fig.~\ref{fig:animation}. The top left panel shows the
phase histories, i.e., the variation of the effective potential value at the true minimum $V_{\rm eff}(v_{h}(T), v_{s}(T))$ as a function of temperature, $T$.
In this plot the temperature drops from right to left. The color shading represents the variation of $m_{h_2}$ values. Not all points on this plane achieve a nucleation temperature $T_n$ and  have a successful phase transition. For those that do, we use a black dot to denote its location on the plane, i.e., the potential and value of $T_n$ at the corresponding vacuum. The magenta 
point denotes the starting benchmark point. The top right panel shows the magenta and black dots on the $g_{211} - m_{h_2}$ plane, while the green dots correspond to points that do not achieve a nucleation temperature $T_n$. The blue band corresponds to the blue points on the right panel of Fig.~\ref{fig:br}, i.e., points with a valid nucleation temperature but any SNR. It is clear that the black dots constitute a single line cutting through the band structure, and from Fig.~\ref{fig:br} (bottom-right panel), we know that moving from the left to the right of this line, we obtain  signals with larger SNR. The  two bottom panels show the black and magenta points on the plane of $(\alpha, m_{h_2})$ and $(\beta/H_n, m_{h_2})$, respectively. It is clear that as one approaches the magenta point, $\alpha$ becomes larger and $\beta/H_n$ smaller, implying a larger SNR. The reason that $\alpha$ increases is due to a delayed transition and thus a more supercooled transition at a lower temperature. For $\beta/H_n$, it becomes smaller, which means slower phase
transition and thus enhancement for gravitational wave production. However, a transition that is too slow would be prevented from completion. This makes larger $m_{h_2}$ infeasible in obtaining a valid nucleation temperature $T_n$. This explains why, as one increases $m_{h_2}$ and overshoots the magenta point in the top right panel, the black dots give way to the green dots. The conclusion is that requiring a sufficiently large SNR narrows down the range of possible masses $m_{h_2}$, explaining the emergence of the narrow band structure in the $g_{211}-m_{h2}$ plane.

\begin{table}[b!]
\begin{tabular}{c|c|c|c|c|c|c}
  \hline 
  \hline
Benchmark & $m_{h_2}$ [GeV]& $\sin\theta$& $v_s$ [GeV]  & $b_3$ [GeV]  & $a_1$ [GeV] & SNR \\  
\hline
A & 825 & 0.20 & 34.6 & 1420 & -1364 & 21.9 \\
B & 1068 & 0.15 & 25.0 & 677 & -1810 & 10.3 \\
\hline
\hline
\end{tabular}
\caption{\label{tab:bench_def} Definitions for the benchmark points illustrated in Fig.~\ref{fig:limits} with the corresponding SNR.}
\end{table}

\section{\label{sec:collider}Probing di-Higgs Blind Spots at the LHC}

In this section, we analyze the collider limits on the heavy Higgs resonance $h_2$, focusing on the blind spot regime. Two benchmark points displaying these features are defined in Table~\ref{tab:bench_def}. They are characterised by depleted $h_2$ branching ratio to di-Higgs $\mathcal{BR}(h_2\rightarrow h_1h_1)$, distinct mixing angles $\sin\theta$, and SNR $>10$. We  explore the complementarity between $h_2\rightarrow h_1h_1$ and $h_2\rightarrow VV$ searches to probe these parameter regimes.

\begin{figure}[t!]
    \centering
     \includegraphics[width=0.325\textwidth]{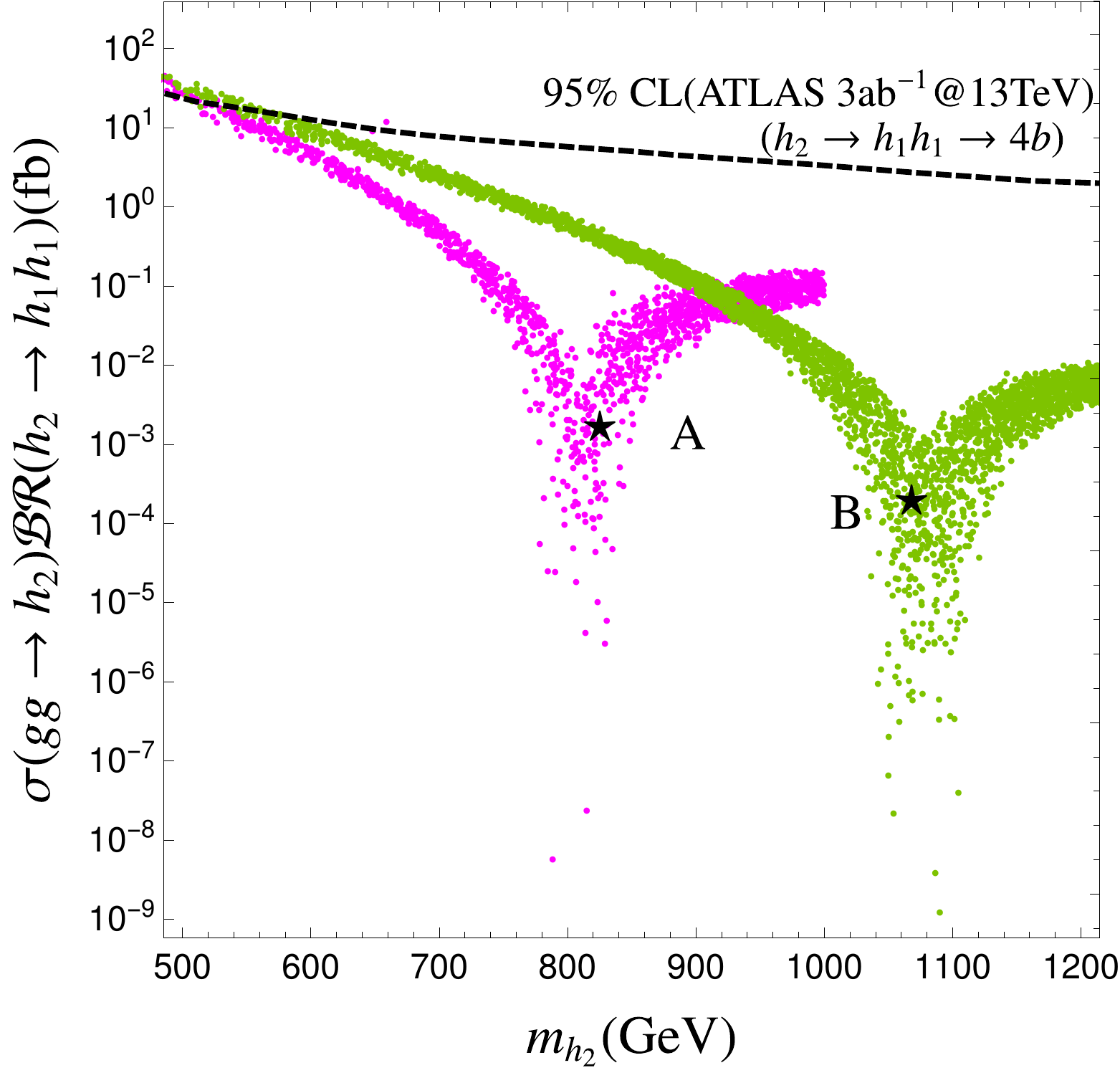}
     \includegraphics[width=0.325\textwidth]{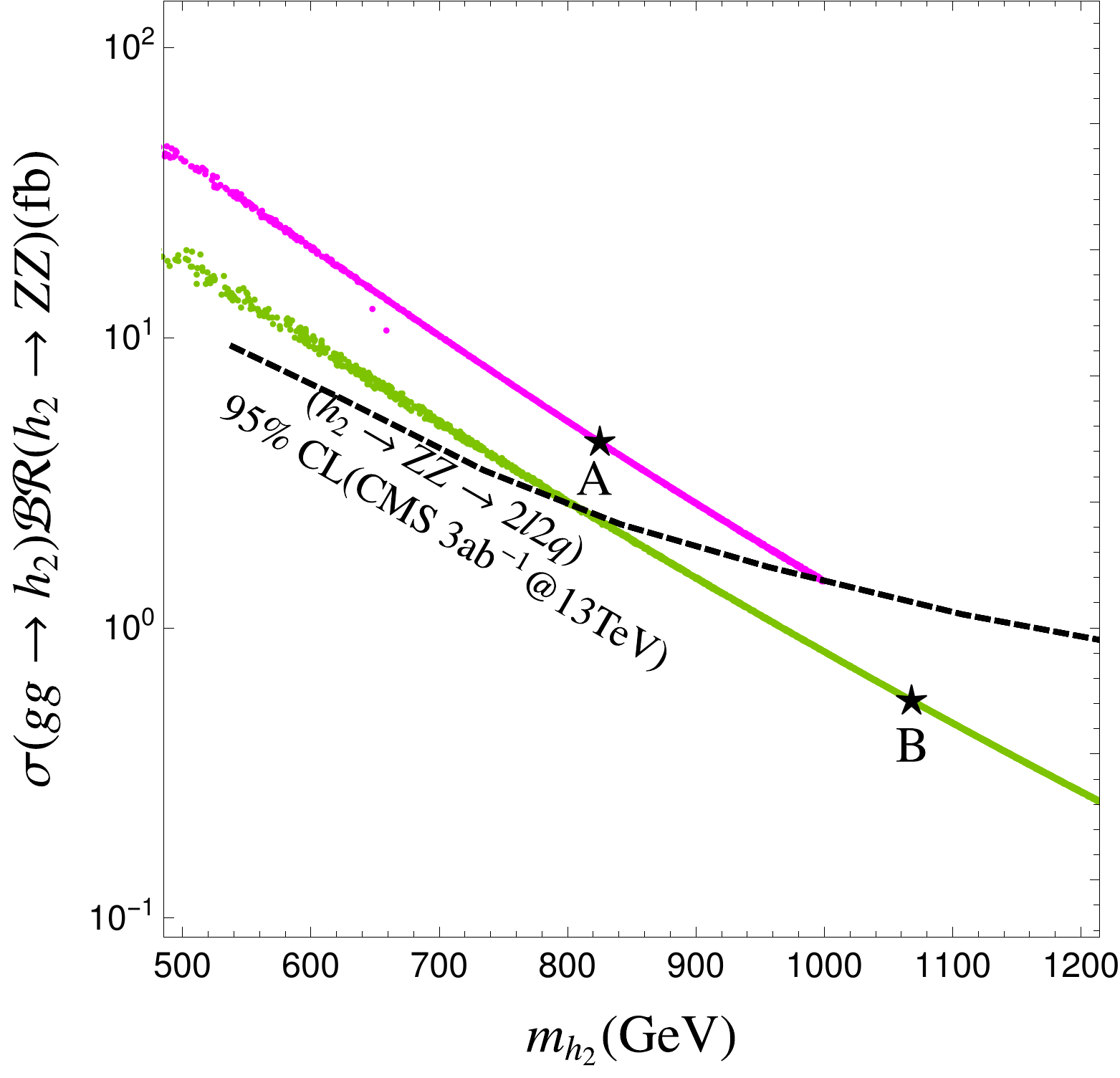}
     \includegraphics[width=0.325\textwidth]{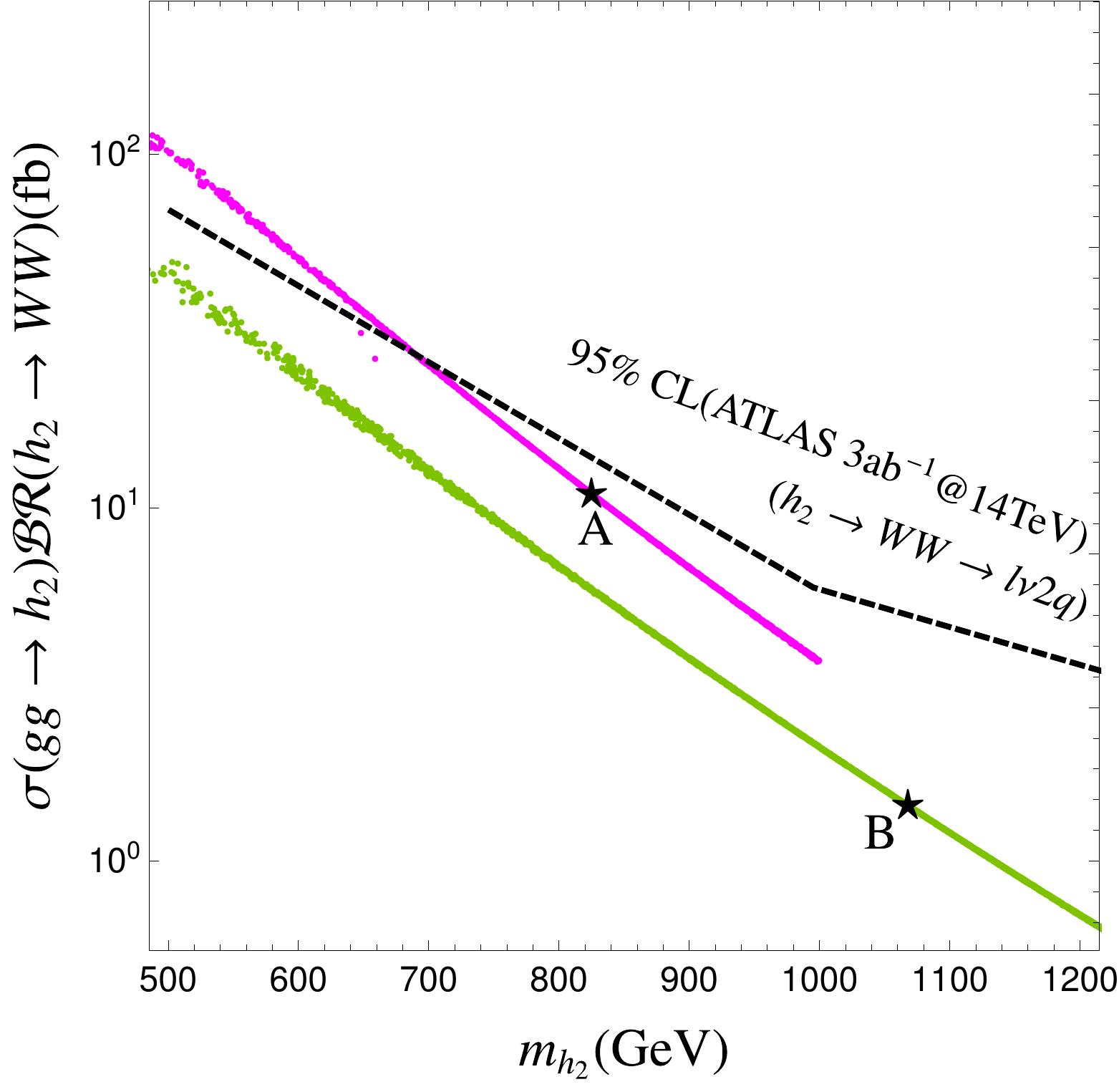}
     \caption{95\% CL limit on the heavy Higgs cross-section decaying to di-Higgs $h_2\rightarrow h_1h_1$ (left panel), di-boson $h_2\rightarrow ZZ$ (central panel), and $h_2\rightarrow WW$ (right panel). The black dashed line represents the LHC limit with 3~ab$^{-1}$ of data. While the $h_2\rightarrow hh$ and $ZZ$ studies assume the LHC at 13~TeV, the $WW$ ATLAS analysis uses the LHC center of mass energy at 14~TeV. The benchmarks A and B, marked as black stars, are defined in Tabs.~\ref{tab:bench_def} and~\ref{tab:bench}. The green and magenta points have $\sin\theta$ fixed at $0.2$ and $0.15$, respectively. All points have SNR $>10$.
     }
    \label{fig:limits}
\end{figure}

We start this phenomenological study focusing on the $pp\rightarrow h_2\rightarrow h_1h_1\rightarrow 4b$ channel. The ATLAS collaboration obtained the current 95\% confidence level limit to this channel in Ref.~\cite{Aaboud:2018knk}. In Fig.~\ref{fig:limits} (left panel), we present the corresponding limit to the heavy Higgs cross-section decaying to di-Higgs (dashed line). The results are scaled to the high-luminosity LHC, $\mathcal{L}=3~\text{ab}^{-1}$. 
  
In addition, we  display  model points that present complementary GWs signals at LISA with $\text{SNR}>10$. For illustration, we show two mixing scenarios: $\sin\theta=0.2$ and 0.15.
The benchmark points defined in Table~\ref{tab:bench_def} are also depicted (black stars). For more details on the respective signal cross-section and branching ratios see Table~\ref{tab:bench}. The  signal cross-section is at NNLO+NNLL QCD and includes top and bottom quark mass effects up to NLO~\cite{deFlorian:2016spz, HXSWG}.   While we optimistically scaled the present ATLAS limits to the high-luminosity LHC scenario without accounting for systematic uncertainties, the benchmarks do not display relevant sensitivities and have rates more than two orders of magnitude below the projected ATLAS constraints. The small branching ratio of the heavy scalar $h_2$ into Higgs bosons $h_1$ results in a large blind spot for the resonant double Higgs searches.

\begin{table}[b]
\begin{tabular}{c|c|c|c|c|c}
  \hline
  \hline
  Benchmark & $\sigma(pp\to h_2)_{13\text{TeV}}$ (fb) & $\sigma(pp\to h_2)_{14\text{TeV}}$ (fb) & $BR(h_2\to ZZ)$ & $BR(h_2\to WW)$  & $BR(h_2\to h_1 h_1)$  \\
\hline
A  & 15.1 & 18.6 & 28.9\% & 58.8\% & 0.0109\% \\
\hline
B  & 1.85 & 2.35 & 30.3\% & 61.2\% & 0.0104\% \\
\hline
\hline
\end{tabular}
\caption{\label{tab:bench} Cross-section and branching ratios associated with the benchmark points A and B defined in table~\ref{tab:bench_def}. The cross-section is at NNLO+NNLL QCD and includes top and bottom quark mass effects up to NLO~\cite{deFlorian:2016spz, HXSWG}.}
\end{table}

\begin{figure}[t!]
    \centering
    \includegraphics[width=0.41\textwidth]{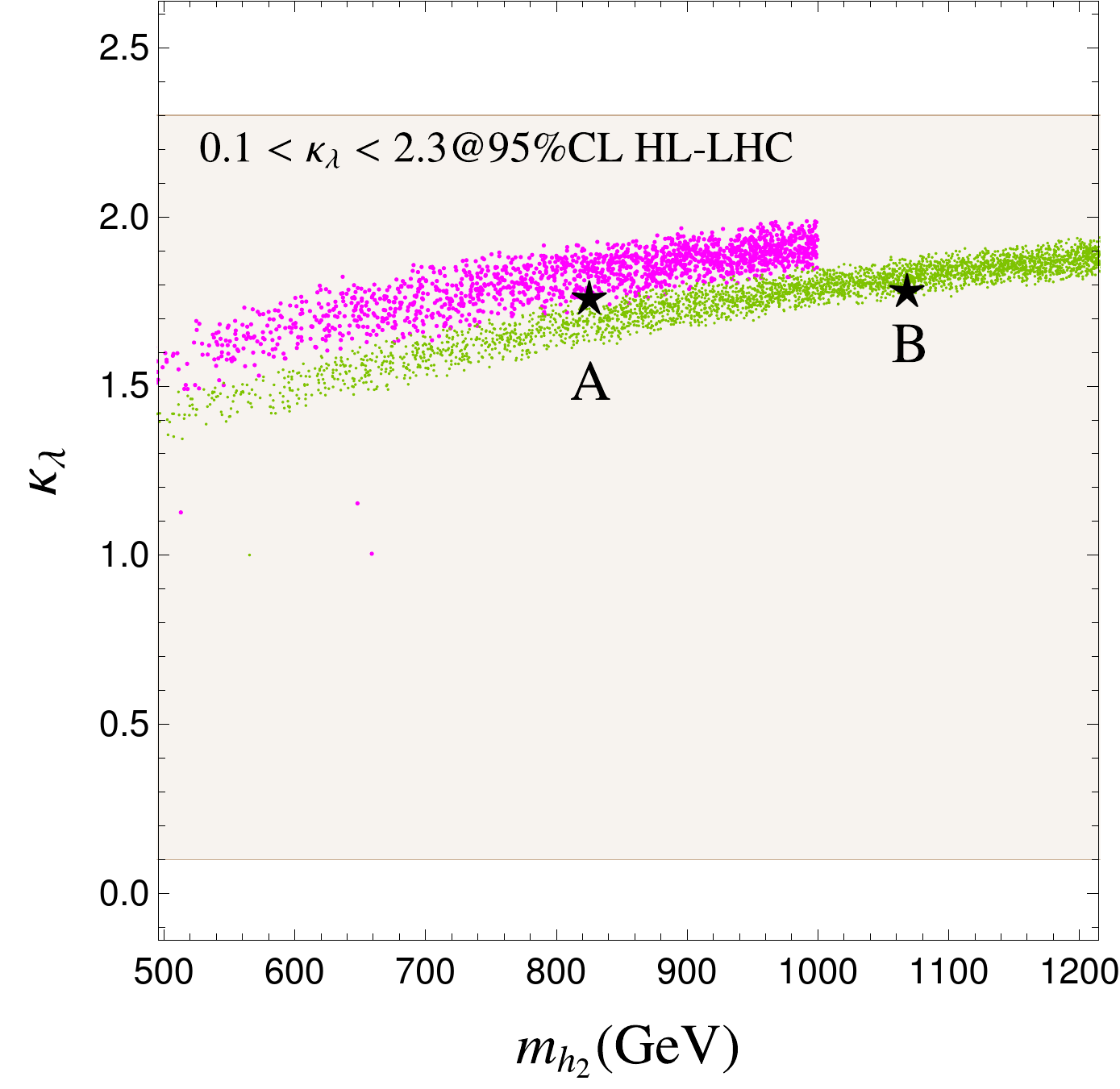}
    \caption{Distribution of parameter space points for GWs signals  in the 
    $(\kappa_{\lambda},m_{h_2})$ plane for $\sin\theta=0.2$ (magenta) and $\sin\theta=0.15$ (green). The projected  95\% confidence level  HL-LHC sensitivity for non-resonant di-Higgs production ${0.1<\kappa_\lambda<2.3}$ is also shown~\cite{Cepeda:2019klc}.
    }
    \label{fig:kappa_lambda}
\end{figure}

In addition to the absence of a resonant peak in the $m_{h_1 h_1}$ distribution for the double-Higgs channel, the  benchmarks considered  have a modified trilinear Higgs couplings $g_{111}$. The interference between the triangle and box diagrams in the non-resonant $h_1h_1$ production is increasingly destructive for $\kappa_{\lambda}$ between 1 and 2.4, where $\sigma(pp\to h_1h_1)$ reaches a minimum. The benchmark points display $\kappa_\lambda\approx 1.8$, resulting in a suppressed $h_1h_1$ cross-section  to approximately half of the SM rate, see Fig.~\ref{fig:kappa_lambda}. Since the current ATLAS and CMS high-luminosity LHC projections indicate that the  trilinear Higgs coupling will be poorly probed $0.1<\kappa_{\lambda}<2.3$ at 95\%~CL~\cite{Cepeda:2019klc}, we should not expect an observation of non-resonant double Higgs production in these blind spot scenarios, in addition to their blindness to the resonant $pp\rightarrow h_2\rightarrow h_1 h_1$ channel.

Now we move on to the complementary di-boson channels $h_2\rightarrow VV$, where $V=Z,W$. The CMS and ATLAS collaborations studied the high-luminosity LHC projected sensitivities to heavy Higgs resonant searches in the channels $pp\rightarrow h_2\rightarrow ZZ\rightarrow 2\ell 2q$ at $\sqrt{S}=13$ TeV and $pp\rightarrow h_2\rightarrow WW\rightarrow \ell\nu 2q$ at $\sqrt{S}=14$ TeV, respectively~\cite{CMS:2019qzn,ATL-PHYS-PUB-2018-022}. The results are shown in Fig.~\ref{fig:limits} (central and right panels). While the di-Higgs searches are blind to the benchmark points defined in Tab.~\ref{tab:bench_def}, the di-boson analyses result in better limits, benefiting from the large heavy Higgs branching ratios to $VV$, see Tab.~\ref{tab:bench}. We observe that the $ZZ$ search will present sensitivity to the di-Higgs blind spot parameter region for mixing $\sin\theta=0.2$. In fact, the bulk of parameter points that lead to GWs signals at LISA with $\sin\theta=0.2$ can  also be probed at the LHC, using the $ZZ$ channel. Notice that  $W$-mass constraint excludes the region with $m_{h2}\gtrsim 1$~TeV for $\sin\theta=0.2$~\cite{Lopez-Val:2014jva,Robens:2015gla}. 
Whereas the smaller mixing scenario $\sin\theta=0.15$ is more challenging at colliders, due to the depleted event rate $\sigma(pp\rightarrow h_2)\propto \sin^2\theta$,  it also  displays relevant phenomenological complementarities between LISA and LHC for $m_{h2}\lesssim 800$~GeV. 

\section{Summary}
\label{sec:summary}

Future gravitational wave experiments, such as LISA, will provide complementary  information to collider experiments on the shape of the Higgs potential. The conditions for strong first-order phase transition and generation of observable GW signals are, however, very restrictive to the profile of the Higgs potential. Using the xSM model as a template, we have shown that the production of signals relevant for future GW experiments can favor feeble $h_2h_1 h_1$ interactions  and characteristic  Higgs self-couplings  in phenomenologically relevant $\sin\theta$ and $m_{h_2}$ regimes. These coupling regimes result in suppressed cross-sections for both resonant and non-resonant di-Higgs signals. While this parameter space is allowed by the theoretical and phenomenological constraints on the model, the restriction to this parameter region is only established after requiring observable GW signals. 

Given the importance for the complementarity picture between GW and collider experiments, we have performed a comprehensive study on the emergence of these di-Higgs blind spot regimes. The requirement for high latent heat release $\alpha$, slow phase  transition (i.e., small $\beta/H_n$), and large SNR induce a clear band structure on the $(g_{211}, m_{h_2})$ plane. This dependence is ultimately driven by the term $a_1 H^\dagger HS$ in the Higgs potential, that controls the size of the tree level barrier in the effective potential, and small $v_s$ with sub-leading dependence on the other free model parameters.

While GWs can favor parameter space regimes resulting in null di-Higgs searches, we show that the complementarity between colliders and GW experiments can be restored  in these parameter regions after  accounting for both di-Higgs and di-boson channels. We perform such an analysis using the high-luminosity LHC projections for resonant $h_2\rightarrow h_1h_1$, $ZZ$, and $WW$ searches. We find that the LHC will be sensitive to the bulk of points displaying GWs signals at LISA with $\sin\theta=0.2$ and to  points with $m_{h2}\lesssim 800$~GeV with $\sin\theta=0.15$.

\begin{acknowledgments}
AA  thanks Conselho Nacional de Desenvolvimento Cient\'{i}fico (CNPq) for its financial support, grant 307265/2017-0. DG was supported by the US  Department  of  Energy under grant number~{DE-SC 0016013}. 
TG was partly supported by US~Department of Energy grant number~DE-SC0010504, U.S.~Department of Energy grant number~DE-FG02-95ER40896, and in part by the PITT-PACC. TG  thanks the Centre for High Energy Physics, Indian Institute of Science Bangalore, where part of the work was performed and Rohini Godbole who made his visit to Bangalore possible.
KS and HG are supported by the U.S. Department of Energy grant number~DE-SC0009956.
\end{acknowledgments}

\bibliographystyle{utphys}
\bibliography{mybib,xsm,4b}
\end{document}